\documentclass[11pt,english]{article}

\pdfoutput=1

\usepackage[a4paper]{geometry}
\usepackage{verbatim}
\usepackage{graphicx}
\usepackage{esint}
\usepackage{colortbl}
\usepackage{color}

\definecolor{light-gray}{gray}{0.85}

\providecommand{\tabularnewline}{\\}

\usepackage{multirow}
\usepackage{babel}

\title{A comparison of methods to determine neuronal phase-response curves}
\author{Ben Torben-Nielsen, Marylka Uusisaari, Klaus M. Stiefel}\date{}

\begin{document}

\maketitle

\begin{abstract}
The phase-response curve (PRC) is an important tool to determine the excitability type of single neurons which reveals consequences for their synchronizing properties. We review five methods to compute the PRC from both model data and experimental data and compare the numerically obtained results from each method. The main difference between the methods lies in the reliability which is influenced by the fluctuations in the spiking data and the number of spikes available for analysis. We discuss the significance of our results and provide guidelines to choose the best method based on the available data. 
\end{abstract}

\section{Introduction} \label{sec:introduction}

The phase-response curve (PRC) of a regularly firing neuron quantifies the shift in the next spike time as a function of the timing of a small perturbation delivered to that neuron. The PRC is an important measure
for several reasons. First, the ability of neurons to synchronize
in excitatory coupled pairs, chains or networks can be predicted from the
PRC: type-I PRCs (purely positive, all excitatory perturbations lead to an acceleration
of spiking) do not allow synchronization while type-II PRCs (biphasic,
acceleration or delay of spiking depending on the phase of the perturbation)
allow synchronization with excitatory connections and short delays \cite{hansel1995}.
Furthermore, the PRC is informative about the type of bifurcation
leading from rest to spiking \cite{izhikevich2007}, thus constraining
quantitative models of the neuron under investigation. Also, the PRC
is correlated with the type of excitability of a neuron \cite{hodgkin1952,marella2008}.

More precisely, a regular firing neuron can be seen as a stable oscillator with period T and only the phase $\phi$ to describe its state. T results from the characteristic angular velocity $\omega$ of the oscillator, thus $\frac{d\phi}{dt}=\omega$. In the absence of inputs a regularly firing neuron fires exactly when $\phi=kT$ (with $k$ an integer and T corresponding the the average ISI, $\widehat{ISI}$). Now suppose an input to that neuron with a small amplitude at phase $\phi$, $\Pi(\phi)$. Then, the influence of this perturbation on the next spike time is described by $\frac{d\phi}{dt}=\omega+\Pi(\phi)Z(\phi)$ where $Z(\phi)$ is the PRC. In other words, the time required to reach the next spike deviates from $T$ according to the perturbation and the PRC. Since the exact spike times, perturbations $\Pi(\phi)$ and $\omega$ (i.e.,$\frac{2\pi}{T}$) are known, we can estimate the PRC $Z(\phi)$ from these data.

Several methods have been proposed to compute the PRC from experimental or modeled data. For basic neuronal models, the PRC can be directly computed from the underlying differential equations by the adjoint method \cite{ermentrout1996}, but for all other cases the PRC has to be determined numerically (for complex models) or experimentally (for real neurons).
In this work, we review five methods for determining PRCs and compare their performance on data sets containing modeled data and experimental data. We identify pitfalls in estimating the PRC, lay out guidelines for approximating the PRC, and assess the reliability of the resulting PRCs.

\section{Methods}\label{sec:methods}

Here we concisely outline five methods to estimate PRCs, describe the
data sets used in the comparison, and describe how we will compare the outcomes. In addition, we describe the direct method which is used to benchmark the performance of the other five methods. In the direct method, the PRC is constructed by injecting excitatory pulses at different phases of the inter-spike interval and measuring the resulting phase shift of the next spike.The PRC is produced directly by plotting the phase of the pulse on the x-axis and the resulting phase shift on the y-axis. In a noise-free case, such as a deterministic simulation, a fine-grained PRC can be generated by injecting pulses at many different phases. 

\subsection{Five methods to determine a PRC}

In the case of experimental data or stochastic simulations, the
data points resulting from applying the direct method will be jittered, and it is necessary to either fit a curve to these points \cite{galan2005,tsubo2007} or bin them \cite{reyes1993,stiefel2009}. We include one such
method in our review of methods (Galan's method, see below).
Due to the often quite significant jitter, it is required to measure
the spike time shift in hundreds of ISIs at randomized
phases. Furthermore, it is necessary to intersperse them with inter-spike
intervals without perturbing pulses to avoid entrainment of spiking
and to have an unperturbed baseline to compare them to. Thus, large
amounts of data are necessary to determine the PRC with the direct method.

\begin{figure}[h!]
\centering
\includegraphics[angle=0,scale=0.6]{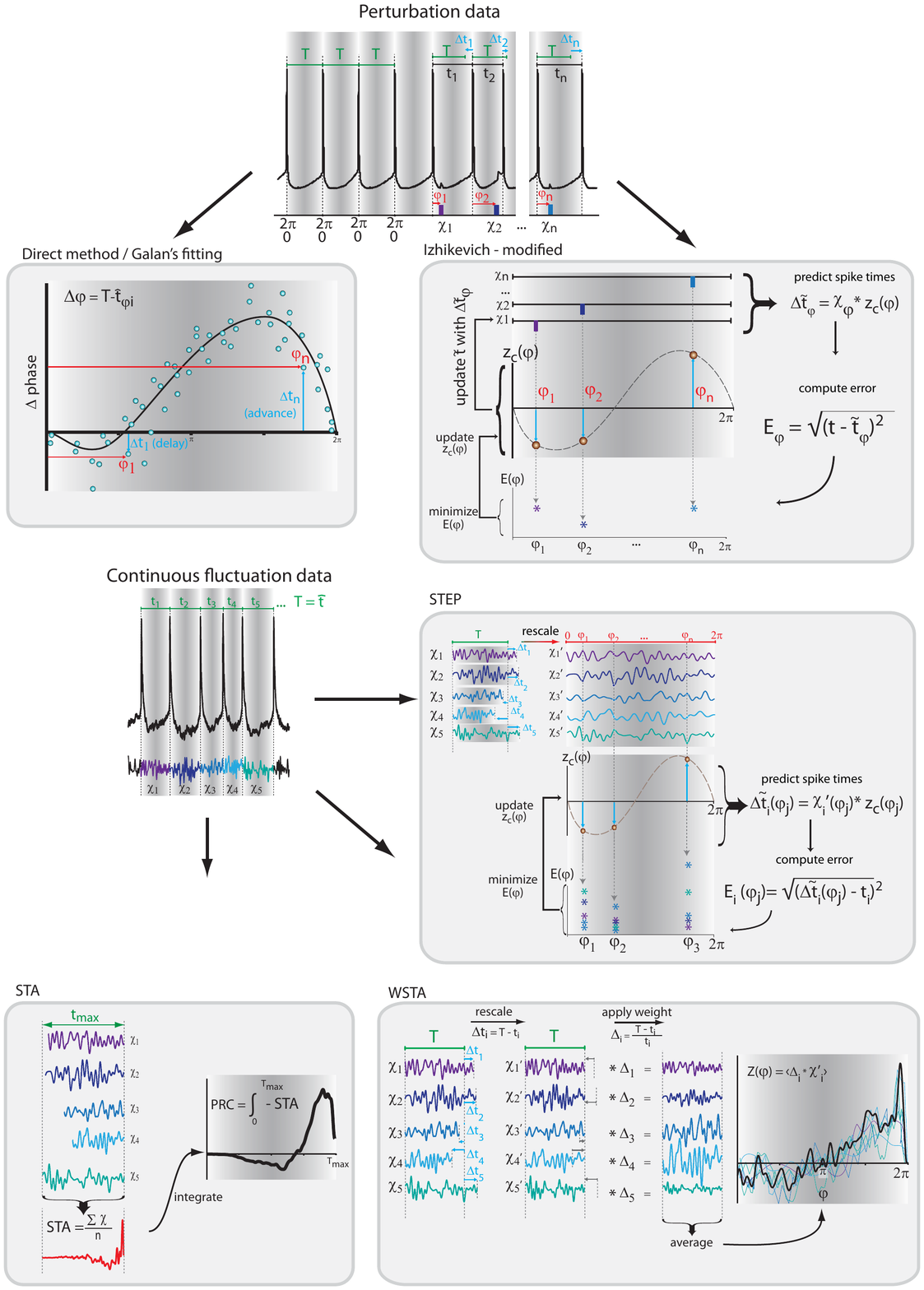}  
\caption{Schematic overview of the reviewed methods. Two methods use pulsed-perturbation data and three methods use continuous fluctuation data. More details in the main text and Table~\ref{table:overview}.}
\label{fig:methods} 
\end{figure}

To alleviate this problem, novel methods have been proposed that use predictions of how spike times will be altered by incoming pulses, and, methods that use continuous fluctuation signals to obtain a more robust PRC measurement based on less spikes. The five reviewed methods consist of one variation of the direct method (Galan's method), two methods that use spike-time predictions to reconstruct the PRC (the modified-Izhikevich method and the STEP method), and two methods that derive the PRC from the incoming continuous fluctuating signal (the STA and WSTA method). The different methods are outlined
below and illustrated in Figure~\ref{fig:methods}.

\subsubsection{Galan's method}

Galan's method \cite{galan2005} uses pulses as perturbations (see the top panel of Figure~\ref{fig:methods}), and fits the PRC to the spike time shifts as a function of the phase of the
perturbation. This is one of the methods which is an extension of
the direct method for noisy data \cite{stiefel2009,tsubo2007,reyes1993}. The PRC $Z(\phi)$ is substituted by a truncated Fourier series, i.e., $Z(\phi)\approx\sum\limits _{0}^{n}{{\bf {a}}\sin(n\phi)+{\bf {b}}\cos(n\phi)}$ and the parameters describing the curve ($\bf{a}$ and $\bf{b}$) are optimized using the Euclidean distance between the data points and the curve as an error signal. The resulting curve is the best approximation of the PRC (only constrained by the length of the expansion of the Fourier series). 

\subsubsection{Modified-Izhikevich method}

Izhikevich proposed an inverse solution to compute the PRC in \cite[chapter10]{izhikevich2007}. The method relies on predicting the next spike time and minimizing the error between the predicted spike time and the true next spike time. The prediction is based on the sum of the phase shifts that a small perturbation (part of a continuous fluctuation) would cause. However, the proposed method does not converge to a correct solution after a reasonable number of fitting rounds (e.g., in about 2-3 hours of computation while other methods converge within minutes). Therefore, we modified the method and used it with less complex perturbation data. In the modified-Izhikevich method,  one pulse $x(t)$ is injected per phase and the next spike is predicted according to a candidate PRC, i.e., $\tilde{s}_{t+1}=x(\phi) z_c(\phi)$ in which $z_c(\phi)$ is the candidate PRC. Then, the candidate PRC is optimized to match the spiking data by computing an error signal proportional to the difference between the predicted next spike time and the actual next spike time, e.g., $Err_{\phi}=\sqrt{\left(s_{n+1}-\tilde{s}_{n+1}\right)}$.

\subsubsection{Spike-triggered average method (STA)}

In the spike triggered average method \cite{ermentrout2007} the the spike-triggered average
is computed from continuous low amplitude current fluctuations (see Figure~\ref{fig:methods}). Then, this spike-triggered average is numerically integrated to produce the PRC. The connection
between the integral of the spike-triggered average and the PRC is
proven for regularly firing neurons and small perturbations
in \cite{ermentrout2007}.  Formally, this method uses the fact that with STA(${\bf {s})}$$=\left\langle {x(s_{n}-s_{n-1})}\right\rangle$
(which is on the interval $[0,ISI_{max}]$ with $ISI_{max}$ being the largest ISI
of ${\bf {s}}$), the relationship $PRC\equiv \int\limits _{0}^{ISI_{max}}{-STA({\bf {s}})}$ holds. In contrast to both Galan's method and the modfied-Izhikevich method, only a single pass over the complete noise signal and the voltage trace is required because there is no optimization step. From this single pass, the spike-triggered average is computed and subsequently integrated. 

\subsubsection{Weighted spike triggered average method (WSTA)}

The weighted spike triggered average method devised in \cite{ota2009}
is an extension of the STA method and also integrates the continuous
low amplitude current fluctuations to derive the PRC. However, in
the WSTA method, the fluctuations in between the different spike times
are normalized to the average ISI ($\widehat{ISI}$) of all spikes in ${\bf {s}}$ ($\widetilde{s_{i}}\equiv\frac{\widehat{ISI}}{\tau_{i}}t$,
with instantaneous ISI $\tau_{i}$). Then, as Ota et
al. \cite{ota2009} prove, with an appropriate weighting function the weighted
sum of these normalized stretches of current fluctuations constitutes
the PRC (for regularly firing neurons). The weighing function is
$\alpha=\frac{{(\widehat{ISI}-\tau_{i})}}{{\tau_{i}}}$. Therefore, PRC $\approx WSTA(\widetilde{{\bf {s}}})\equiv\left\langle {\alpha x(\widetilde{s_{i}})}\right\rangle $.

\subsubsection{Standardized error prediction method (STEP)}

The STEP method \cite{torben200X} is an extension of the modified-Izhikevich
method to work with continuous fluctuation data (in a different way than originally proposed by Izhikevich). In this method, instead of using a prediction error averaged over all ISIs and all phases, the temporal information in the error is preserved by binning the errors of all ISIs independently per phase.

The fluctuations are binned equidistantly on $\widehat{ISI}$ (i.e., normalized) and are treated as if they were independent, i.e., for each phase bin of each ISI, the predicted next spike time is computed and the mismatch with the true next spike time is used to optimize the parameters of a curve representing the PRC. More precisely, all inter-spike intervals are normalized to $[0,2\pi]$ and
discretized into N bins. For each bin, an independent prediction
is made about the next spike time: $\tilde{s}_{i,j}=\omega+\left[{x(bin_{j})z_c(\phi)}\right]$, where $\phi$ corresponds to phase of  $bin_j$.
Subsequently, one obtains a 2-dimensional array with the the dimensions
given by N bins and M spikes. Finally, a least-squares fitting algorithm
is used to minimize the 2-D prediction error array and obtain a PRC
that predicts the recorded spike time shifts \cite{torben200X}.

Table~\ref{table:overview} summarizes the five implemented methods (plus the direct method) and how they
relate to each other. In addition to the reviewed methods, there are several other published methods which we omitted because they proved impractical (with respect to experimental demands), e.g., the MAP-estimation algorithm \cite{ota2009b}
and the post-stimulus time histogram method \cite{gutkin2005}.

\begin{table}[htb]
\begin{center}
\begin{tabular}{|l|l|l|}
\hline 
 & \textbf{No optimization}  & \textbf{Optimization}\tabularnewline
\hline 
\multirow{2}{*}{\textbf{Perturbation data}}  & (Direct method)  & Galan's method\tabularnewline
 &  & Izhikevich-derived method\tabularnewline
\hline 
\multirow{2}{*}{\textbf{Continuous fluctuation data}}  & STA  & STEP \tabularnewline
 & WSTA  & \tabularnewline
\hline
\end{tabular}
\caption{Comparison of the implemented methods in terms of the required data and the requirement to optimize the outcome.}
\label{table:overview} 
\end{center}
\end{table}

\subsection{Data sets}\label{subsec:data}

We tested the five methods (and the direct method) with three different data
sets whenever possible. The first two data sets contain model data while the third set contains experimental data. We used the single-compartmental model as developed by Golomb and Amitai \cite{golomb1997} and modified by
\cite{stiefel2009} to generate the data. This model uses a Hodgkin-Huxley-type formalism to model
neural spiking behavior:

\[
C_M \frac{{dV}}{{dt}} =  - m^3 h\overline g _{Na} \left( {V - E_{Na} } \right) - n\overline g _{KDR} \left( {V - E_K } \right) - s\overline g _{Ks} \left( {V - E_K } \right) - \overline g _{leak} \left( {V - E_{leak} } \right) - I_{inj} 
\]
and $\frac{{dx}}{{dt}} = \tau \left( V \right)\left( {x - x_\infty  \left( V \right)} \right)$, where $V$ is the membrane potential, $\overline{g}_{x}$ the maximum conductance for ion $x$ and $E_x$ the reversal potential for ion $x$. The parameter values can be found in \cite{golomb1997,stiefel2009}. By turning the adaptation current on or off, this model switches between
type-II or type-I excitability \cite{stiefel2009}, respectively.

\begin{figure}[htb]
\begin{center}
\includegraphics[scale=0.7]{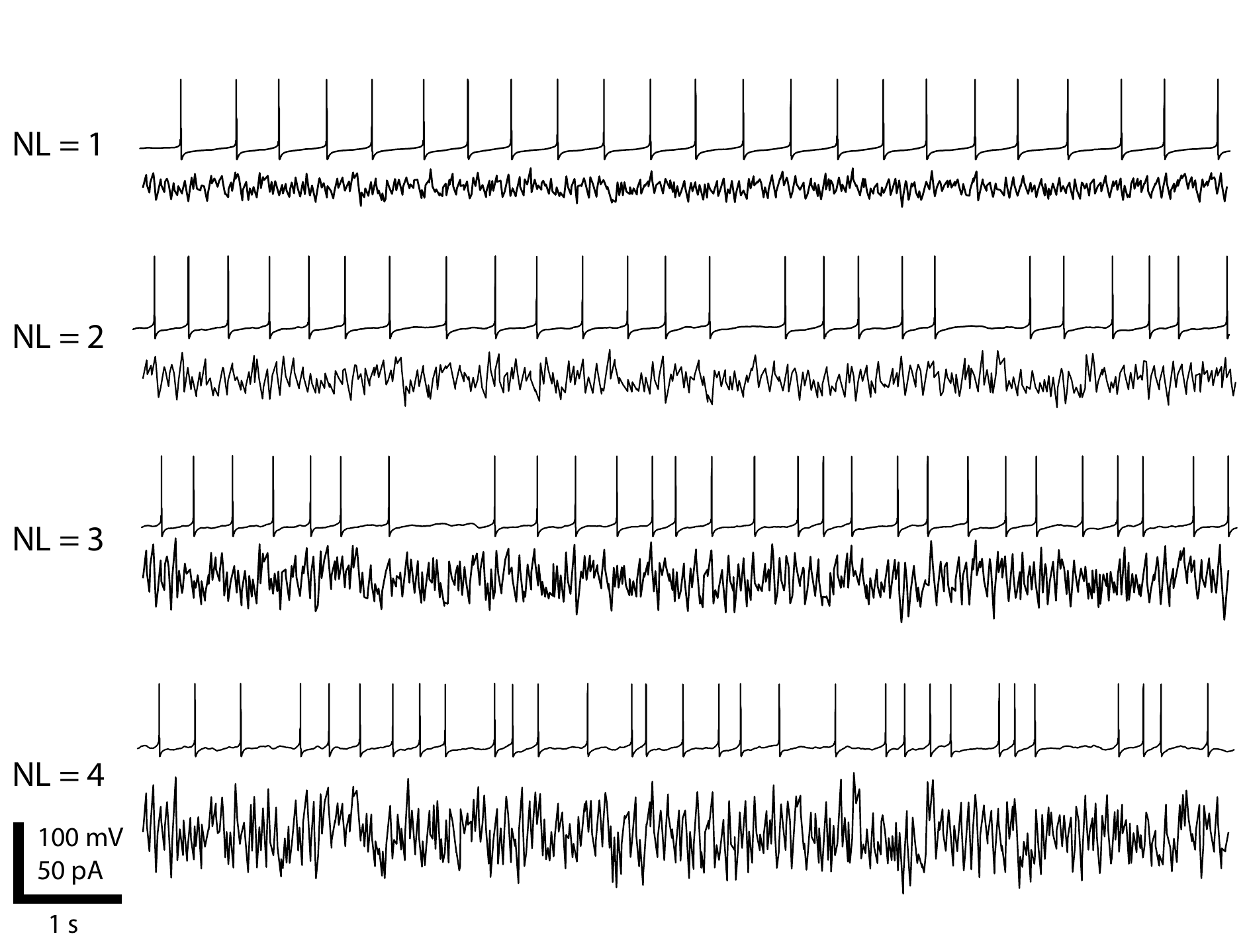}
\caption{Noise and the resulting spike trains as generated by our model neurons. The four noise levels have increasing amplitudes around the same mean and are generated by an Ornstein-Uhlenbeck process. The noise has a profound influence on the regularity of the spikes.}
\label{fig:noise}
\end{center}
\end{figure}

The first data set contains noise-free model data in which a
single compartmental model neuron (see below) is perturbed at different
phases. This set contains 128 pulses evenly spaced over $[0,2\pi]$.
The second data set contains modeled data from the same single-compartmental
model but with an additionally injected fluctuating current. The fluctuations are
generated through a stationary Orstein-Uhlenbeck process around a
given mean value and parametrized by the reversion rate ($g=0.1$)
and 4 different volatility levels ($D={1e^{-4},5e^{-4},1e^{-5},5e^{-5}}$).
The advantage of the noisy modeled data is that the
excitability type is known with certainty because small perturbations do not
change the PRC type \cite{izhikevich2007}.  The injected fluctuating current and the resulting spike trains are illustrated in Figure~\ref{fig:noise}.  The different noise levels result in four groups of data each containing approximately 950 spikes. The noise and the resulting spike trains are illustrated in Figure~\ref{fig:noise}. The last data set contains experimental data recorded from a layer 3/4 pyramidal cell of the mouse visual cortex with the whole cell patch-clamp
technique in vitro. Standard patch-clamp techniques as in \cite{stiefel2008} were used. Membrane potential voltage data and the injected fluctuations were digitized at 40 kHz, and two levels of current ($\mu=50$ and $100$ pA) were injected as fluctuations. The fluctuations consisted of white-noise low-pass filtered at 200 Hz. In both the model data and experimental data, the fluctuations are on top of a step current ($I_s$) which is required to get the model/cell into a regime of regular firing

\subsection{Quantitative comparison}

To investigate which method produces the most reliable result, we compared the different methods with the calibrated PRC resulting from the direct method. The difference resulting from the five implemented methods with the directly observed PRC can be quantified by examining the Euclidean distance (by taking the mean-squared error, MSE) and correlation (with the Pearson correlation) between the obtained PRCs.

It is important to note that Galan's method and the modified-Izhikevich method cannot be used with continuous fluctuating data because they are designed to work solely with pulsed perturbation data. However, the methods intended for continuous fluctuation data (e.g., STA, WSTA and STEP) can be used to compute the PRC from both perturbation data and fluctuation data because the former is a simplified case of the latter: instead of a continuous stream of fluctuations only a single fluctuation per period is injected. Hence, we can compare the PRCs resulting from the five implemented method with the directly observed PRC on the perturbation data, but only the STA, WSTA and STEP on the more complex continuous fluctuation data (i.e., noisy model data set and experimental data set).

To compare the methods with each other, we normalize the PRCs in a post-processing step. All but the STA method are defined (along the x-axis) over $[0,2\pi]$. We scale the time of the STA produced PRC to the same interval. Along the y-axis we normalize the PRC so that the (positive) peaks are equal to 1. Since most researchers are primarily interested in the type of the PRC (type-I vs. type-II), and because the normalization does not affect qualitative features of the PRC such as the slope and the positive and negative areas, the normalization of the results is valid.

\subsection{Implementation details}

We implemented the different methods in Python in combination with the Numpy/Scipy and Matplotlib\footnote{The code is developed for scientific use and can be obtained from \texttt{http://www.irp.oist.jp/tenu/btn/Tools.html}}. In three methods (Galan's method, modified-Izhikevich and STEP) a curve is optimized to fit the data. Any smooth curve such as a polynomial or a Fourier series can be used for this purpose. To be consistent with the implementation in previously published studies \cite{galan2005,izhikevich2007} we use the third expansion of the Fourier series (n=3) in the remainder of this manuscript. Larger expansion would provide better fits in some cases (when there is a steep slope in the PRC) but it would also be more prone to overfitting and hence the third expansion seems suitable. Moreover, Galan's method and (the original) Izhikevich method do not prescribe a particular optimization algorithm although Galan uses least-squares optimization. We follow his work and also use least-squares optimization in Galan's method, the modified-Izhikevich method and STEP method.

For the WSTA method, the authors suggest to fit a polynomial to the raw outcome of their algorithm because 
this raw output is noisy with a smaller number of spikes. For the sake of clarity we show the raw outcome to illustrate the true capabilities of this method.

All methods require configuration of the estimated inter-spike interval ($\widetilde{ISI} $\footnote{In theory, the average inter-spike interval $\widehat{ISI}$ is required. However, in most cases when firing is not regular and the ISI histogram mildly skewed, we have to estimate the `average' inter-spike interval $\widetilde{ISI}$. In the remainder of this manuscript, $\widehat{ISI}$ and $\widetilde{ISI}$ are used as synonyms.}). In our implementation of the different methods, the $\widetilde{ISI}$ can be given as an argument to the algorithm or automatically computed. The automatic computation straightforwardly takes the mean and, therefore, works only for highly regular firing neurons. In addition we exclude ISIs that do not satisfy $0.1 \times \widetilde{ISI} \leq \widetilde{ISI} \leq 2 \times \widetilde{ISI}$ because larger spread of ISIs generally causes the methods to fail (remember that the PRC is a characteristic of regularly firing neurons.).

\section{Results}\label{sec:results}

\subsection{Performance on noise-free model perturbation data}

\begin{figure}[htbp]
\centering \begin{tabular}{cc}
\includegraphics[scale=0.4]{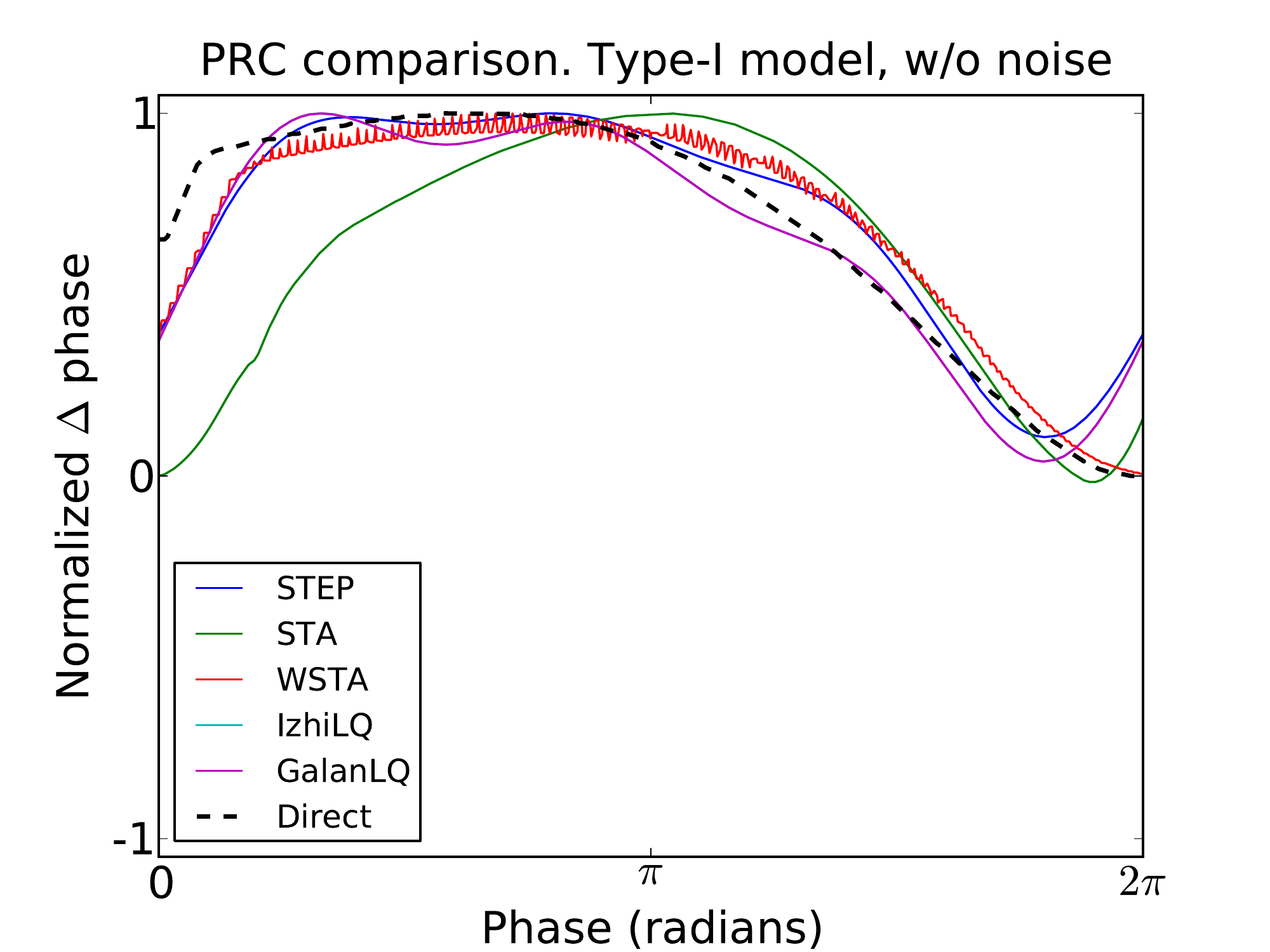}  &
 \includegraphics[scale=0.4]{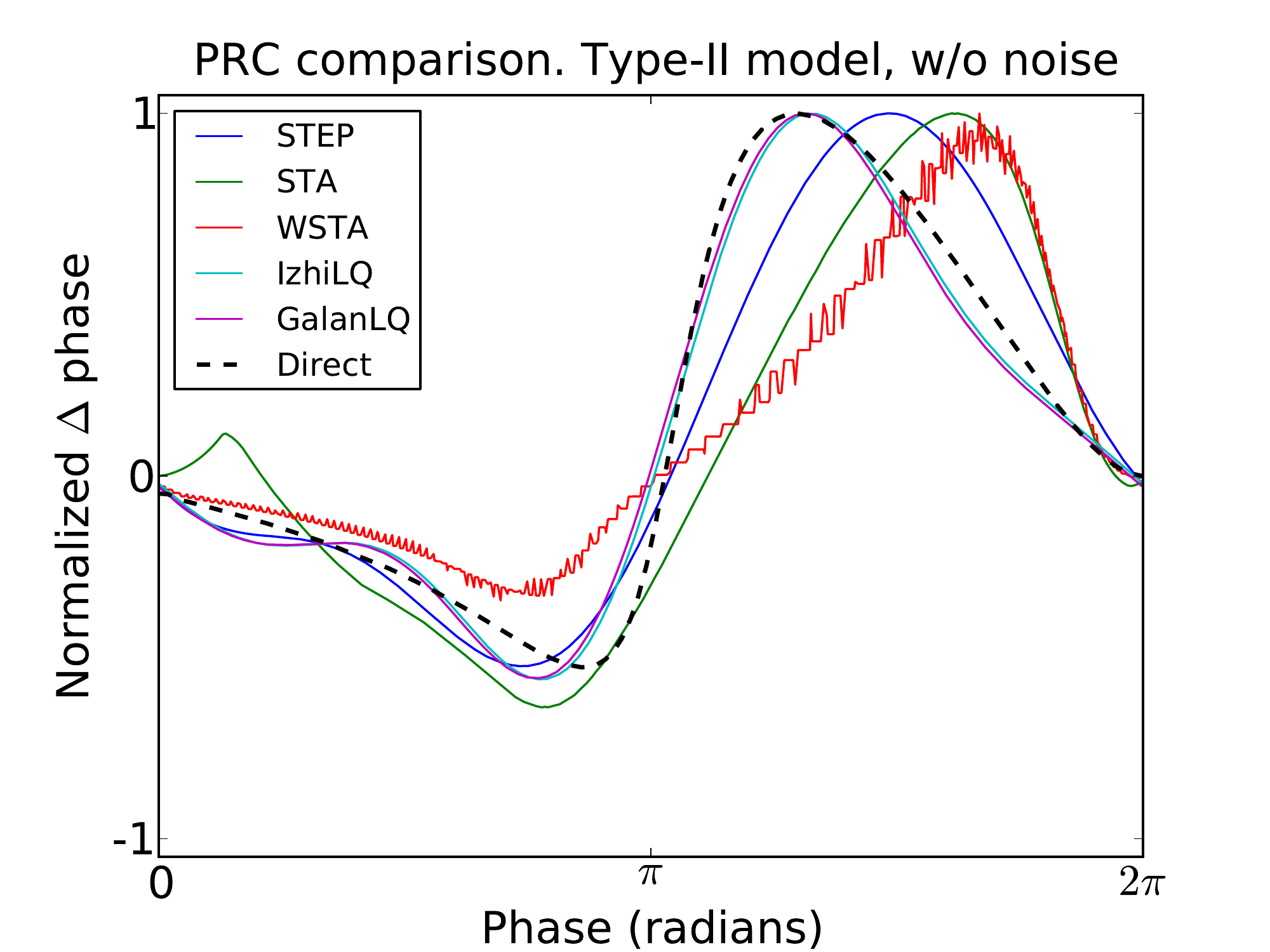}
 \tabularnewline
\end{tabular}
\caption{Comparison of all PRC estimation methods on noise-free model data. The dotted black line is the directly observed PRC. The PRCs from all five methods agree on the PRC type, have a similar shape, and resemble the directly determined PRC.}
\label{fig:direct_compare} 
\end{figure}

Figure~\ref{fig:direct_compare}
illustrates the PRCs  as computed by all implemented methods for noise-free model data with both type-I and type-II parameters. Table~\ref{table:direct} quantifies the difference between each PRC and the directly observed PRC by means of the MSE and Pearson correlation. For brevity, the modified-Izhikevich method is labeled as `IzhiLQ' and Galan's method is referred to as `GalanLQ'; in both cases the LQ suffix indicates the use of least-squares optimization. The PRCs resulting from the direct method is plotted as a dashed line and serves to calibrate the results. It can easily be verified from Figure~\ref{fig:direct_compare} that the five methods compare qualitatively; it can be verified from Table~\ref{table:direct} that they are also quantitatively similar. 

The best results using this type of data are obtained by the methods designed to work with pulse-perturbation data only, namely Galan's method and the modified-Izhikevich method. The results of these two methods are better in terms of the MSE and the Pearson correlation compared to the results of the methods intended for continuous fluctuating data. Moreover, the quantitative analysis shows equal results of the modified-Izhikevich and Galan's method on type-I data. This result is due to the simplicity of the curve to fit and the low dimension of the search space (i.e., 3 values for $\bf{a}$ and $\bf{b}$ of the Fourier series). On type-II data,  the modified-Izhikevich method performs best. From methods intended for continuous fluctuating data, the STEP method has the best performance on both the type-I and type-II model data set as the MSE and Pearson correlation indicate closest resemblance to the directly obtained PRC. 

\begin{table}
\begin{center}
\begin{tabular}{|l|l|l|l|l|}
\hline
\multicolumn{1}{|c|}{\textbf{Method}} &  \multicolumn{2}{c|}{\textbf{Type-I}} & \multicolumn{2}{c|}{\textbf{Type-II}} \tabularnewline \hline
& MSE & Pearson & MSE & Pearson \\ \hline
\textbf{STA} & 0.691571 & 199.188386 &0.833662 &218.912789 \\
\textbf{STEP} &0.951353  &  67.878615 & 0.939346& 130.261712\\
\textbf{WSTA} &  0.959762 & 76.051482 & 0.760755& 235.848697\\
\textbf{IzhiLQ}  & \cellcolor{light-gray}0.961458 & \cellcolor{light-gray}66.999838 & \cellcolor{light-gray}0.991354&\cellcolor{light-gray}54.368821 \\
\textbf{GalanLQ}  & \cellcolor{light-gray}0.961458 &\cellcolor{light-gray}66.999838  & 0.988336&59.415021 \\ \hline
\end{tabular}
\caption{Quantitative performance of the different methods to estimate the PRC. The means-squared error (MSE) and the Pearson correlation with respect to the directly observed PRC are used for the quantification. For both type-I and type-II  data, the modified-Izhikevich method performs best both in terms of MSE and Pearson correlation.  On the type-I data set, the modified-Izhikevich and Galan's method obtain the same performance (and hence perform equal) due to the low degree of freedom (6 parameters to be optimized for the Fourier series). Of the methods designed to work with continuous data, the STEP method performs the best.}
\label{table:direct} 
\end{center}
\end{table}

\subsection{Performance on noisy model data}

\begin{figure}[htbp]
 \centering \begin{tabular}{cc}
\includegraphics[scale=0.4]{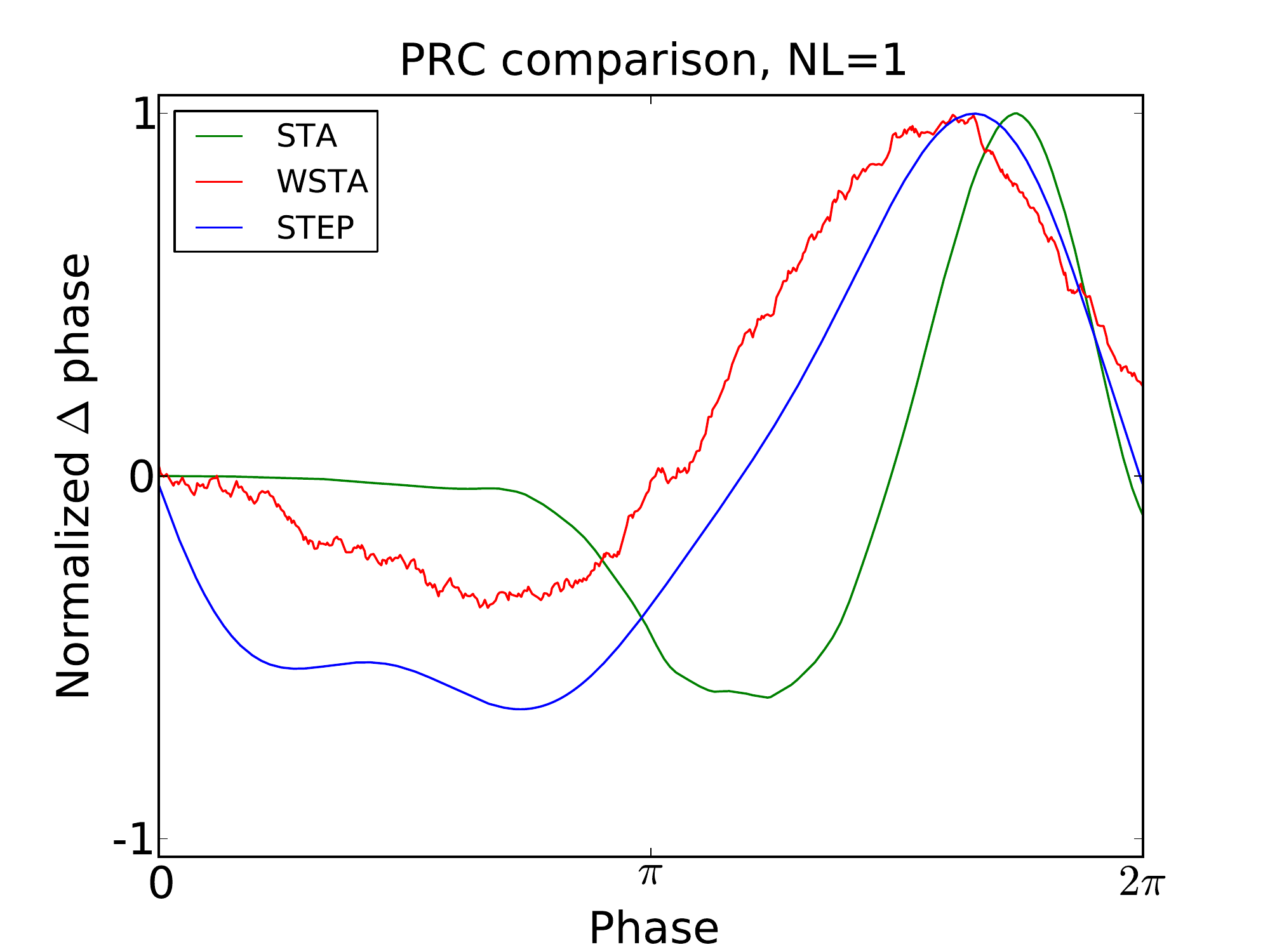}  & 
\includegraphics[scale=0.4]{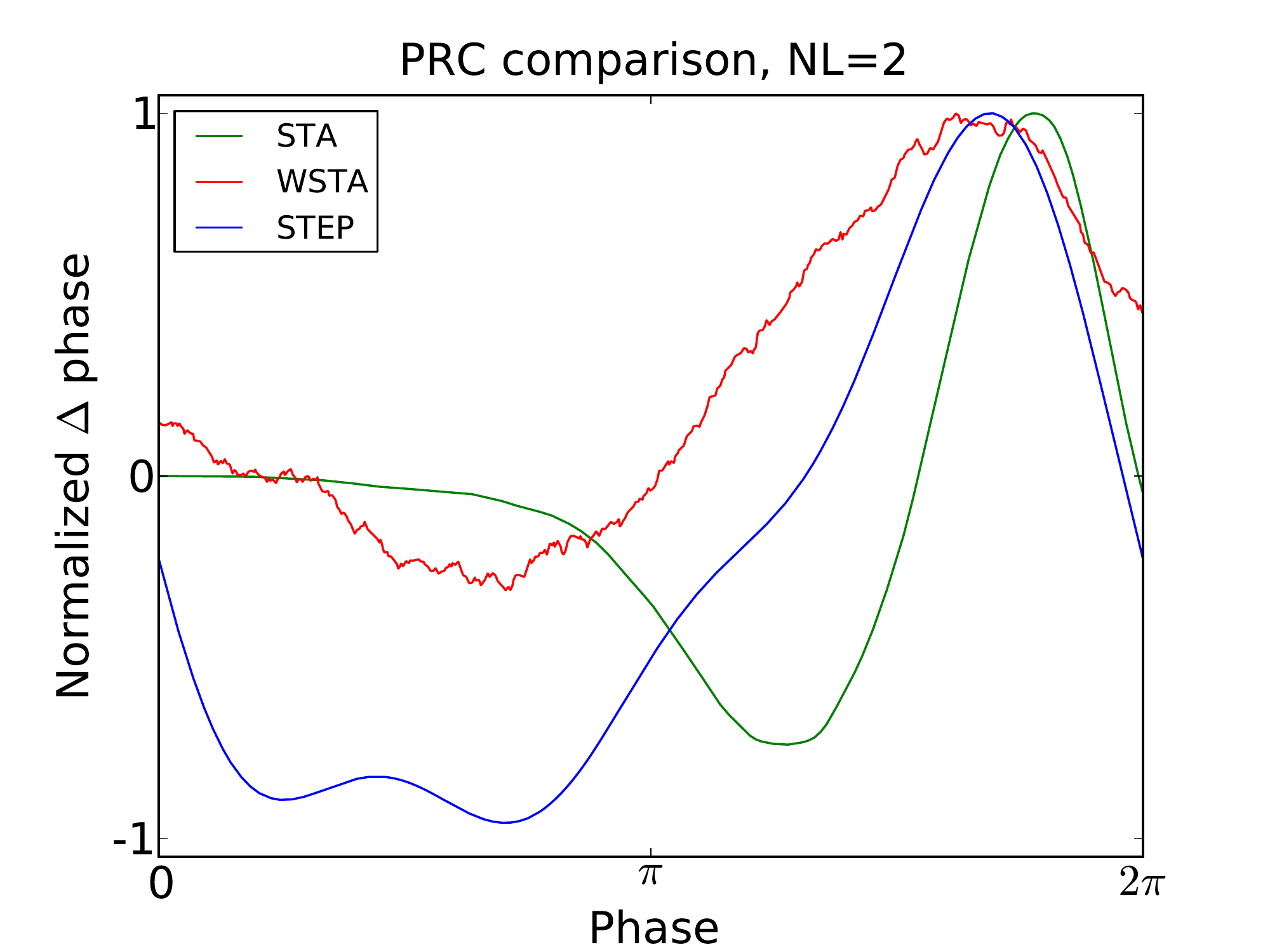} \tabularnewline
\includegraphics[scale=0.4]{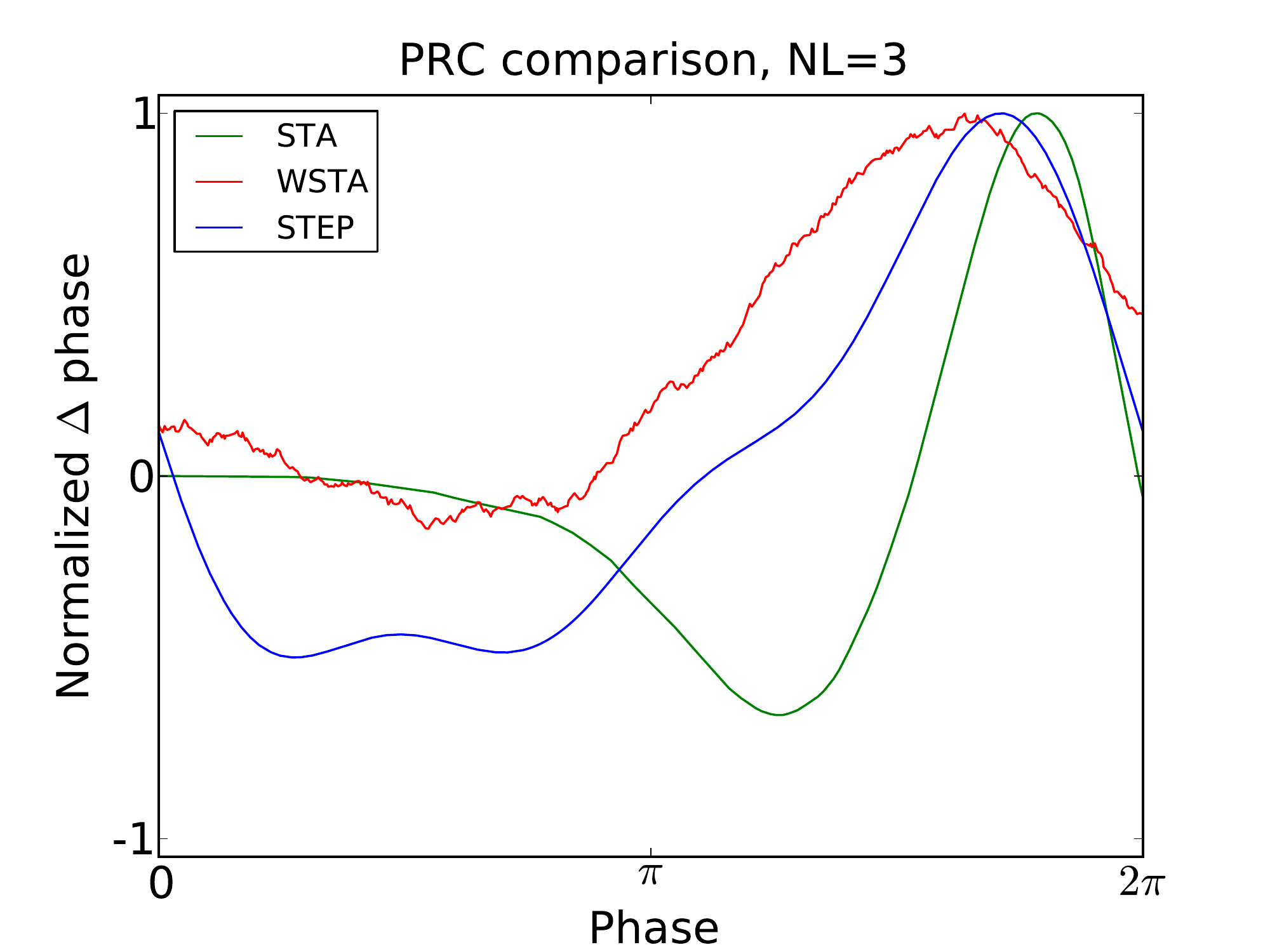}  & 
\includegraphics[scale=0.4]{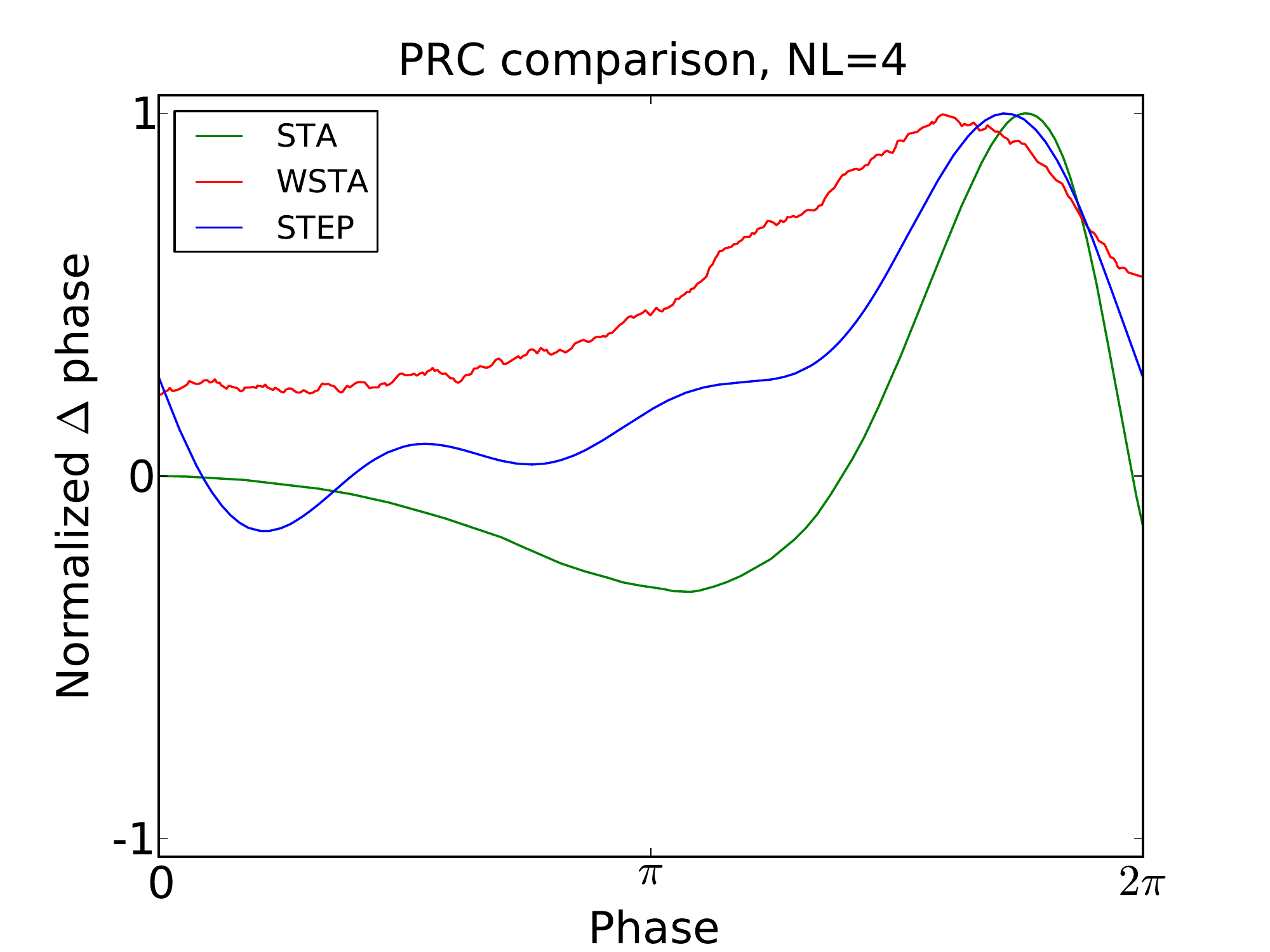} \tabularnewline
\end{tabular}
\caption{Comparison of PRC methods on noisy model data. The four panels illustrate the resulting PRCs at different noise levels (NL=1,2,3,4).}
\label{fig:noisy_compare} 
\end{figure}

The noisy model data are obtained by continuous fluctuating current
injection. Here we compare the three methods designed to analyze such data.
We used four different noise levels for the comparison. Each noise
level has the same mean and only differs in the variance around the
mean \footnote{The reversion rate in the Orstein-Uhlenbeck noise was kept constant
while the volatility was increased.}. Figure~\ref{fig:noisy_compare} illustrates the PRCs produced by the STA, WSTA and STEP method at the four levels of fluctuations. From this figure it is clear that for
low noise levels all methods agree qualitatively on the
PRC type as all methods correctly indicate type-II PRC in the model. However,
for the two higher noise levels, the WSTA method results in a (mostly) nonnegative, type-I, PRC while the two other methods correctly assess the neuron as type-II. Hence, we can say that the STA and STEP method cope better with high amplitude fluctuations. Table~\ref{table:noisy} quantifies the difference between the computed PRCs and the directly observed PRC.  In contrast to the qualitative observation that STA and STEP perform better with fluctuation, the WSTA method has the highest resemblance to the directly observed PRC in terms of MSE and Pearson correlation 
(except at the highest fluctuation level where the STEP method obtains the best Pearson correlation). We explain this observation as follows: with higher amplitude fluctuations, the regularity of the spikes decreases. As a result, the PRC as computed from this less regularly firing data is different than the PRC from the noise-free data \cite{tateno2007,tsubo2007}. However, the PRC is a quantitative measurement of regular firing neurons and hence, the true PRC of a neuron is the PRC measured from spikes in the regular firing regime. Therefore, we compare the PRCs always to the PRC directly observed in the noise-free case although the shape of the PRC at higher fluctuation levels might look different. The PRC produced by the WSTA method has more resemblance to the directly observed PRC although is does provide a wrong categorization of the PRC type at higher amplitudes. For low-amplitude fluctuations we conclude that the WSTA method produces the most reliable PRC, but for higher amplitude-fluctuations (NL=3,4) the STEP and especially the STA method produce more reliable results (Figure~\ref{fig:noisy_compare})).

The reliability of the STA method has to be inferred from a visual inspection of the produced PRC and knowledge of the noise-free, directly observed PRC because the quantitative analysis always indicates low similarity between the PRC produced by the STA method and the noise-free, directly observed PRC. This phenomenon is due to the fact  that the STA method is computed on the interval $[0,ISI_{max}]$ and only afterwards scaled to $[0,\widetilde{ISI}]$. As a consequence, the STA is always relatively flat at the beginning of the normalized interval and the PRC is shifted to later phases. Even though the interpretation of the PRC is beyond the scope of this review, the shift to higher phases does not matter for the reliability as the shift is consistent and proportional to increasing irregularity of the spike times.

\begin{table}
\begin{center}
\begin{tabular}{|l|l|l|l|l|l|l|l|l|}
\hline
\multicolumn{1}{|c|}{\textbf{Method}} &  \multicolumn{2}{c|}{\textbf{NL=1}} & \multicolumn{2}{c|}{\textbf{NL=2}} &  \multicolumn{2}{c|}{\textbf{NL=3}} &  \multicolumn{2}{c|}{\textbf{NL=4}} \\  \hline
& MSE & PC & MSE & PC & MSE & PC & MSE & PC \\ \hline
STA &-0.089941 & 514 &-0.252694 & 562  & -0.191698& 541&0.279241 & 410\\
STEP &0.751355 &  343 & 0.702422&  544& 0.671676& 336 &0.551901 &   \cellcolor{light-gray}387\\
WSTA &\cellcolor{light-gray}0.847633 &  \cellcolor{light-gray}212  & \cellcolor{light-gray}0.776861 &\cellcolor{light-gray}282  & \cellcolor{light-gray}0.815145 & \cellcolor{light-gray}318 & \cellcolor{light-gray}0.788307 & 459\\ \hline
\end{tabular}
\caption{Quantitative performance of the different methods to estimate the PRC using noisy, continuously fluctuating data. The WSTA performs the best on most noise levels in terms of both the MSE and Pearson correlation (PC) when compared to the directly observed PRC. An interpretation of these results is in the main text.}
\label{table:noisy}
\end{center}
\end{table}

\subsection{Performance on experimental data}

\begin{figure}[htbp]
 \centering \begin{tabular}{cc}
\includegraphics[scale=0.4]{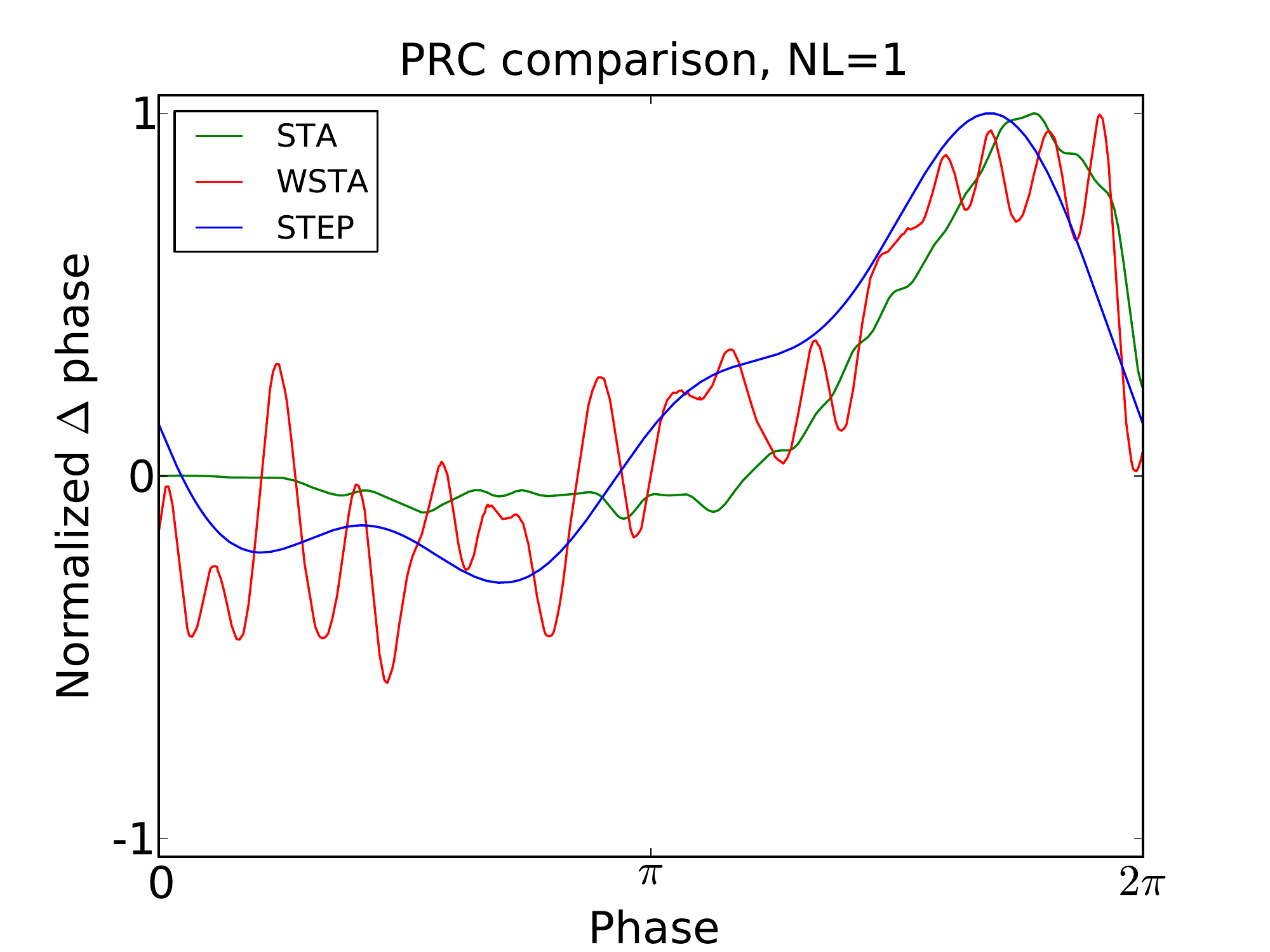}  & 
\includegraphics[scale=0.4]{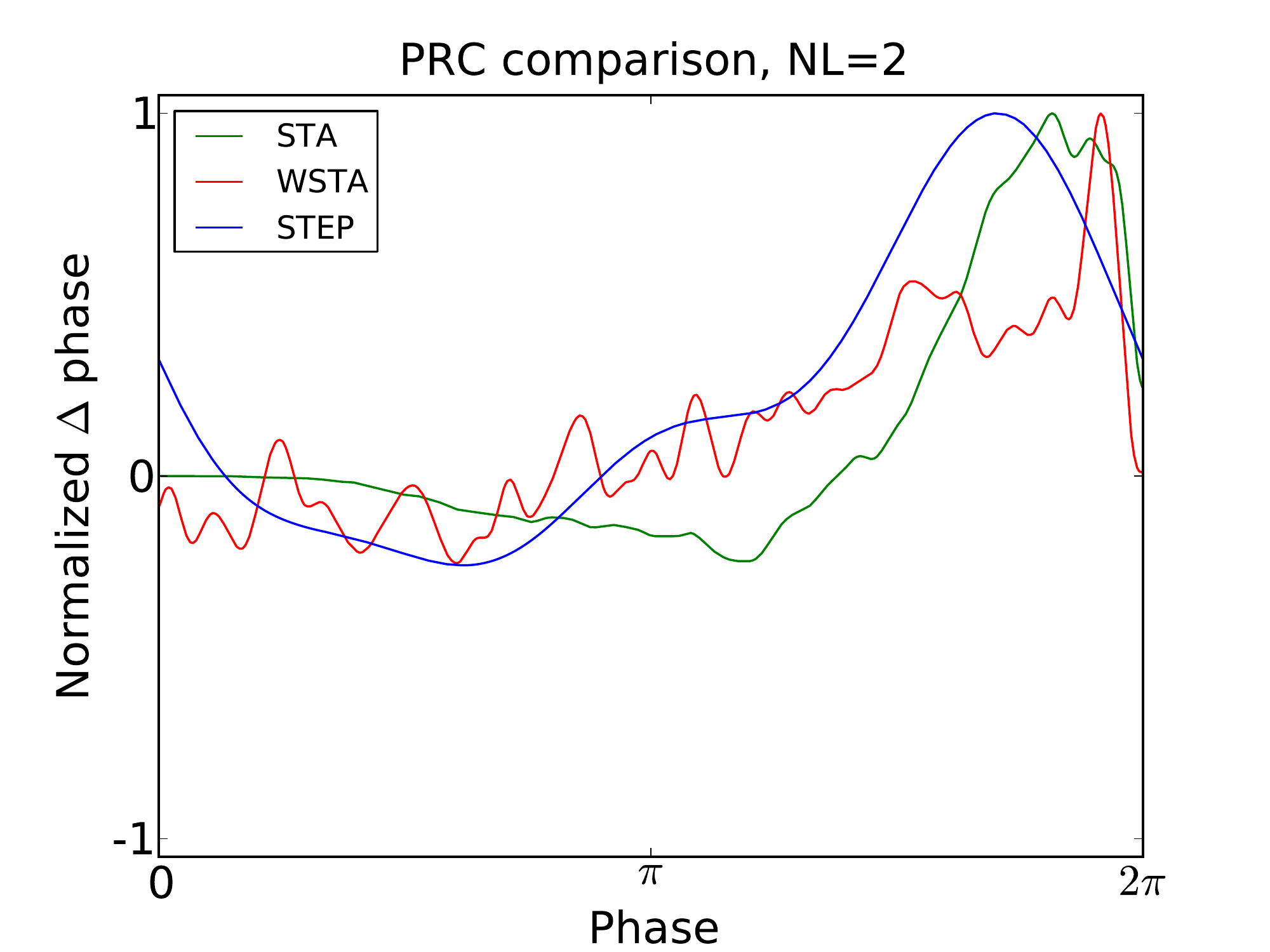} \tabularnewline
\end{tabular}
\caption{Comparison of different PRC methods on experimental data. Two noise levels were tested and the three methods show a fair agreement on the type (type-II) of the PRC curve.}
\label{fig:exp_compare} 
\end{figure}

The most interesting test case is the comparison of different PRC methods applied to real, experimental data. We compare the STA, WSTA and
STEP method on data from layer 2/3 neurons. Figure~\ref{fig:exp_compare}
illustrates the results with two different noise levels. With 650 spikes and considerable spread of ISIs, the WSTA method produces a noisy outcome while the STA and STEP result in smooth PRCs (as they are optimized Fourier series of small expansion). The two noise levels produce PRCs that very similar; all three methods indicate a type-II PRC although at both noise levels the WSTA produced PRC is rather noisy. In the case of experimental data, there is no calibrated data to test against and hence we only look at the Pearson correlations between the different methods to assess the level of agreement between the different methods (Table~\ref{table:experimental}). For the lower noise level, the Pearson correlation between the three methods is always higher than 0.85, indicating good correspondence between the three methods. Also, for the higher noise level, the correlations stay above 0.73 which still indicates good agreement between the different methods. The STEP method has the highest Pearson correlation with the other two outcomes and can therefore be seen as a sort of `average' of the other two methods. Additionally, the high resemblance between the three methods corroborates that these PRCs are reliable.

\begin{table}
\begin{center}
\begin{tabular}{|l|l|l|l|}
\hline
& \textbf{STA} &\textbf{STEP}  & \textbf{WSTA} \\ \hline
\multicolumn{4}{|c|}{Noise 1} \\ \hline
\textbf{STA} & 1.000000 &0.872780& 0.852972\\
\textbf{STEP}& &1.000000 &0.887956 \\
\textbf{WSTA}& & & 1.000000 \\ \hline
\multicolumn{4}{|c|}{Noise 2} \\ \hline
\textbf{STA} &1.000000 &0.794771& 0.737648\\
\textbf{STEP}& &1.000000 & 0.837448 \\
\textbf{WSTA}& & & 1.000000 \\ \hline
\end{tabular}
\end{center}
\caption{Agreement between different PRCs on the experimental data. Shown are the Pearson correlation between the PRCs generated by three different methods. All show strong correlation indicating similar trends in the three curves.}
\label{table:experimental}
\end{table}

\subsection{Reliability and parameter sensitivity}

Here we address the reliability of the various methods for estimating the PRC. Moreover, we investigate how the reliability is affected by parameter sensitivity in the algorithm.

\begin{figure}[htbp]
 \centering \begin{tabular}{ccc}
\includegraphics[scale=0.25]{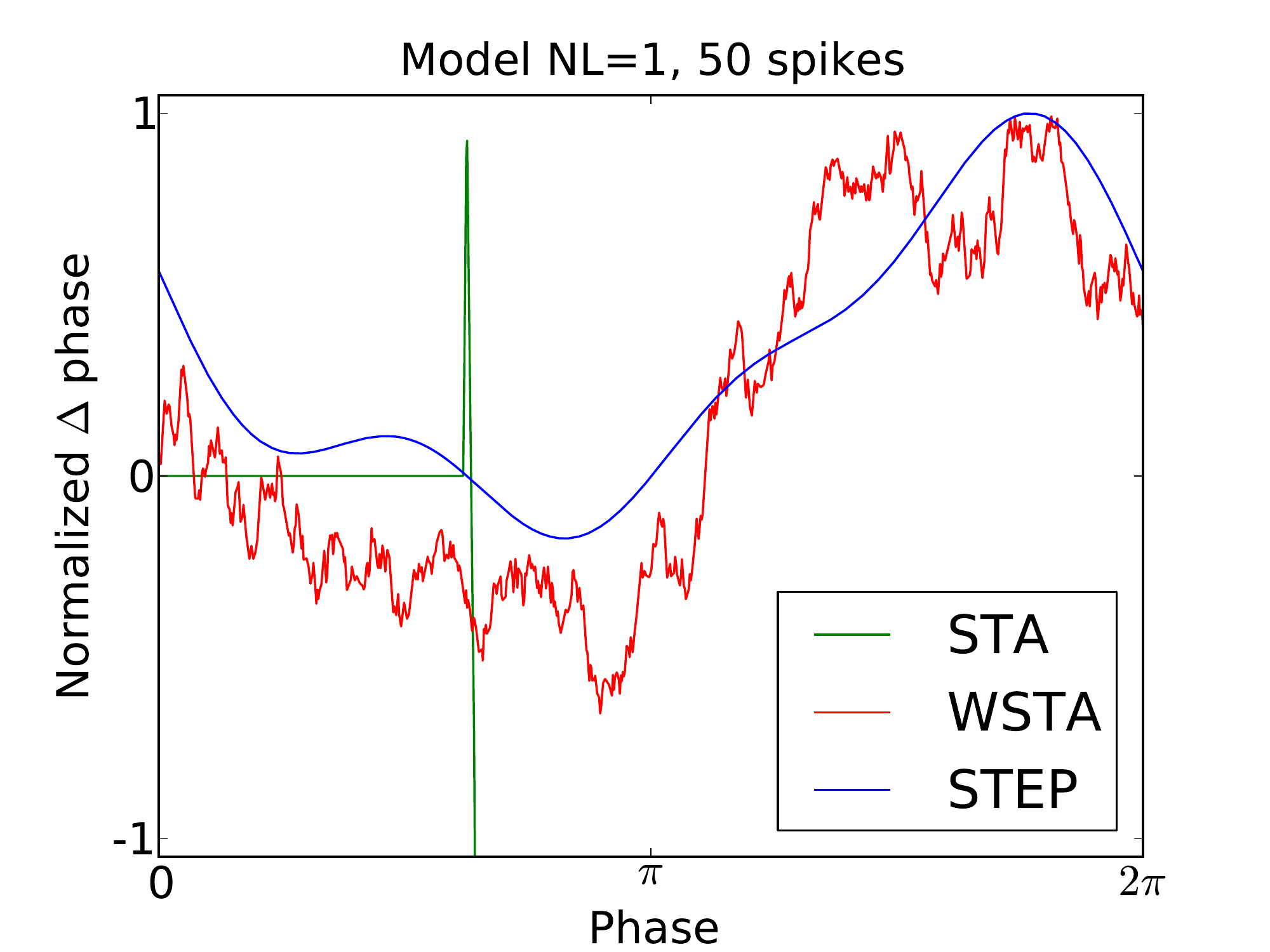}  &
 \includegraphics[scale=0.25]{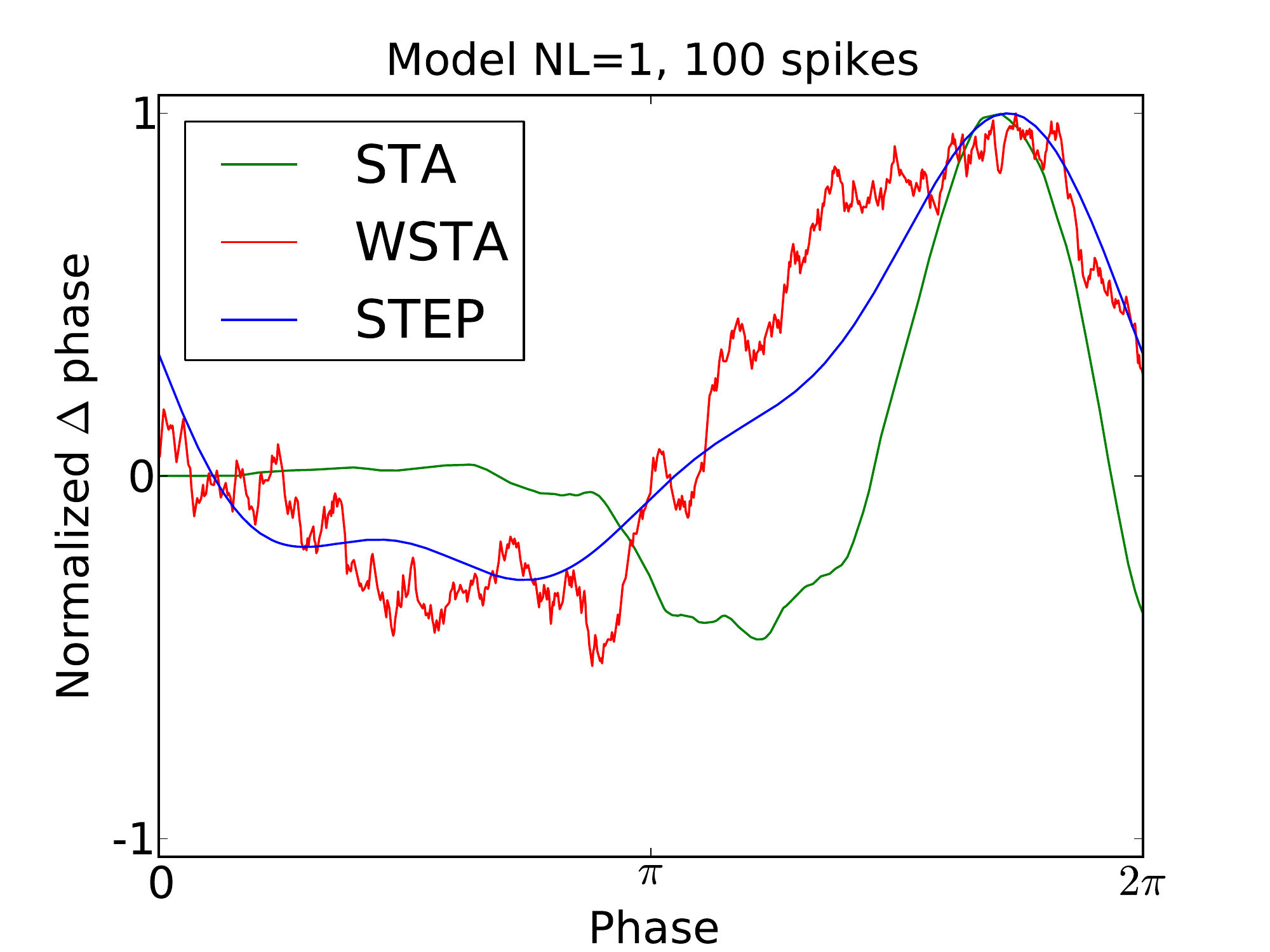}  & 
 \includegraphics[scale=0.25]{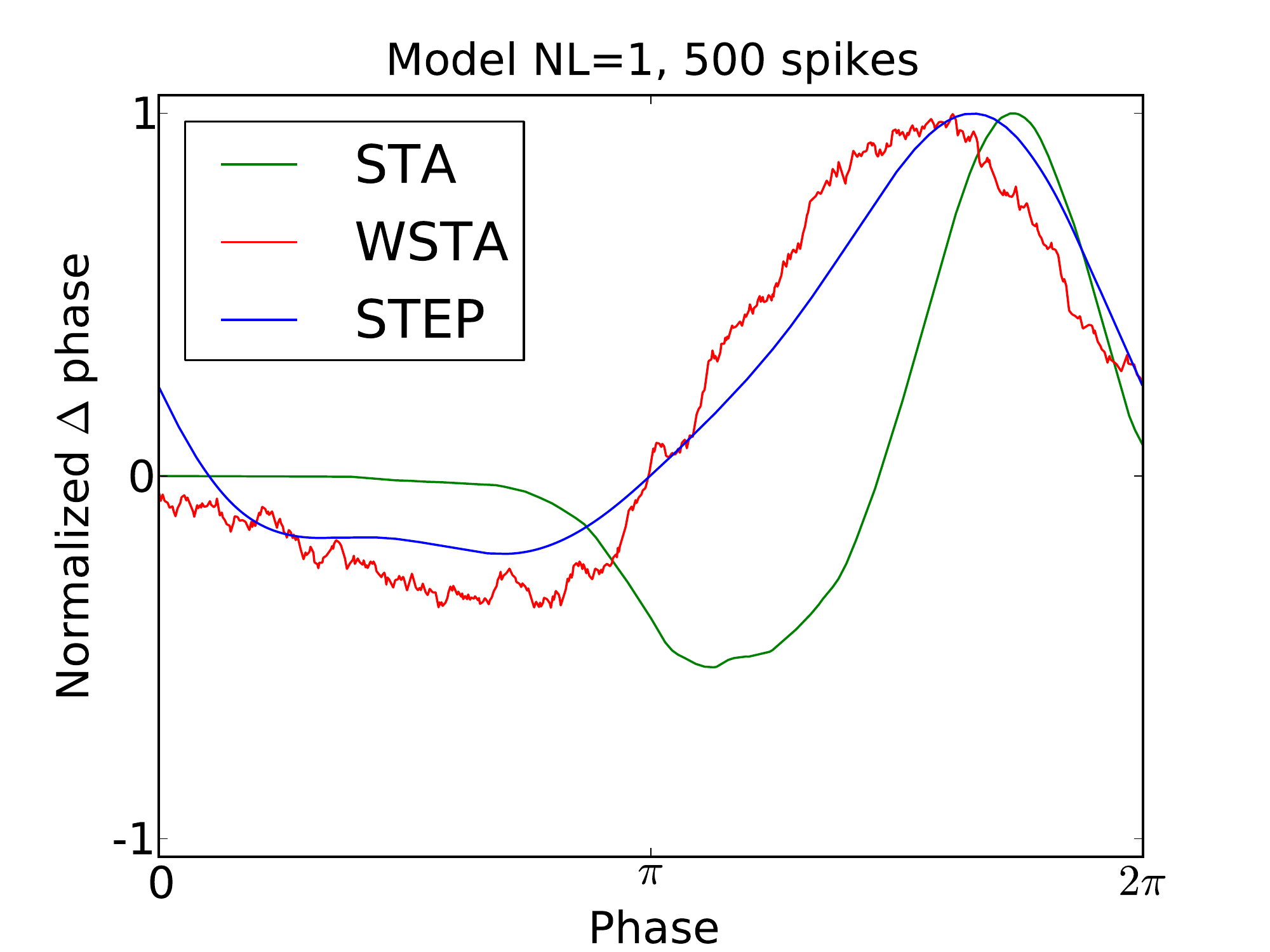} \tabularnewline
\includegraphics[scale=0.25]{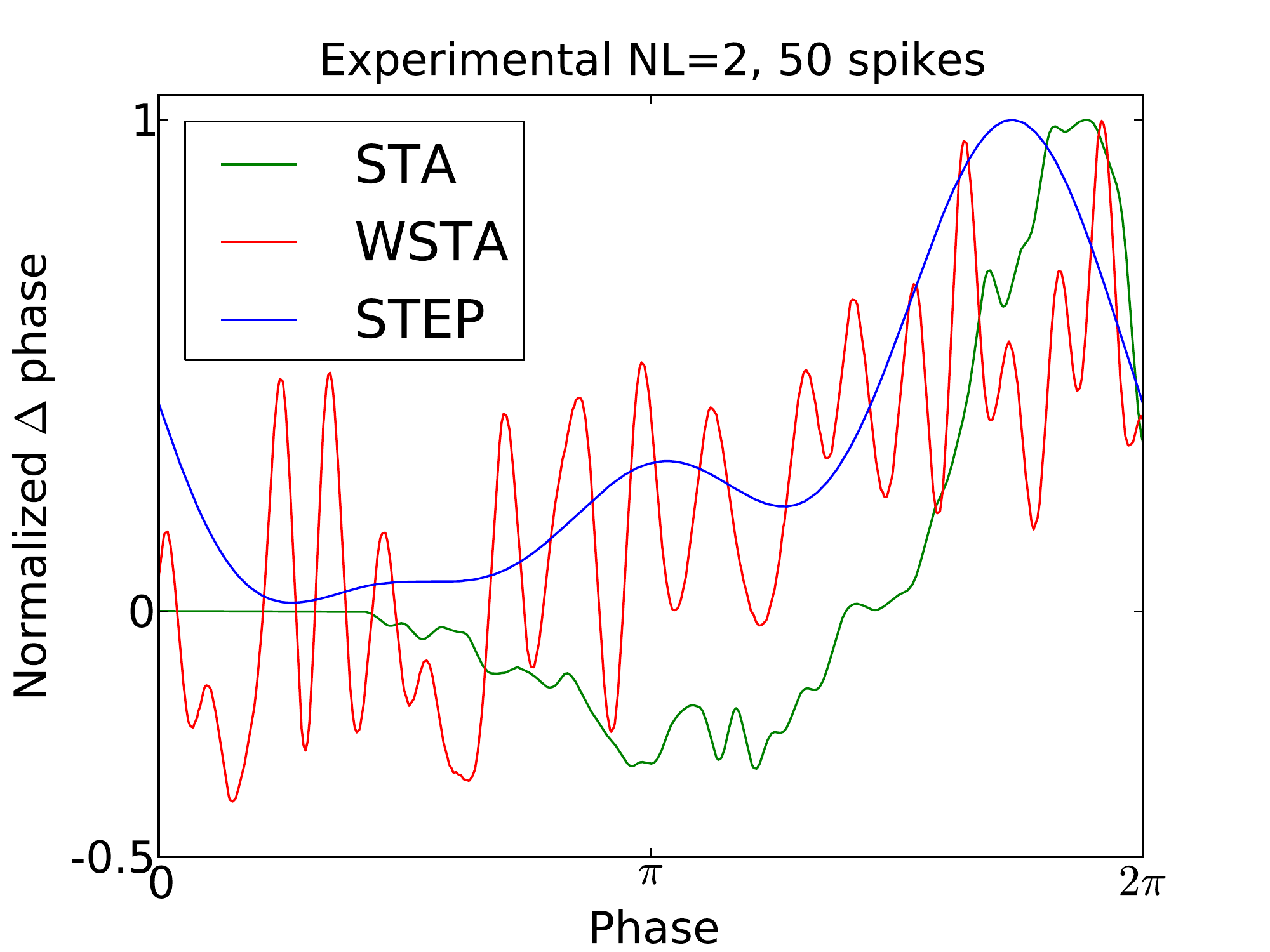}  & 
\includegraphics[scale=0.25]{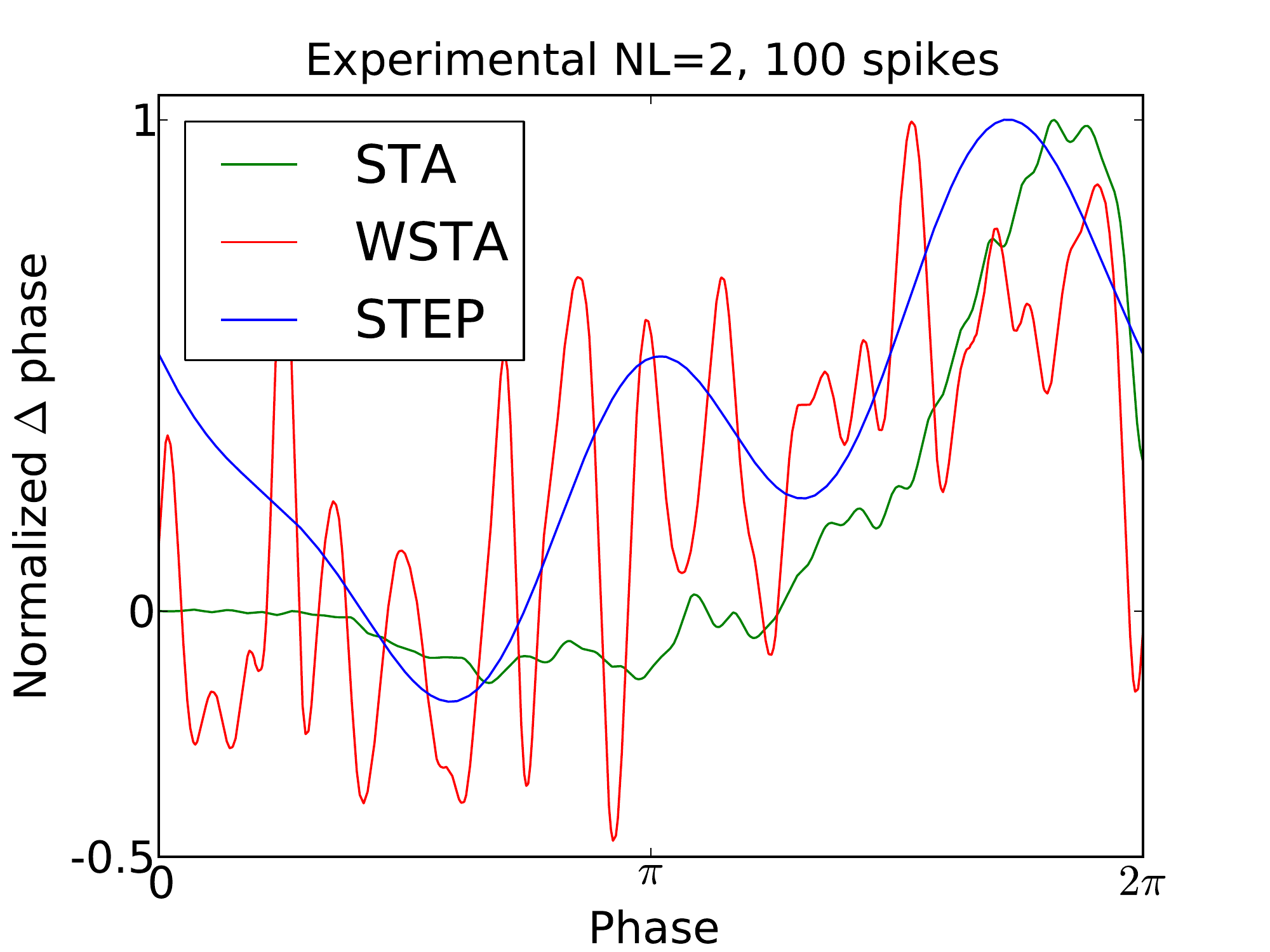}  &
 \includegraphics[scale=0.25]{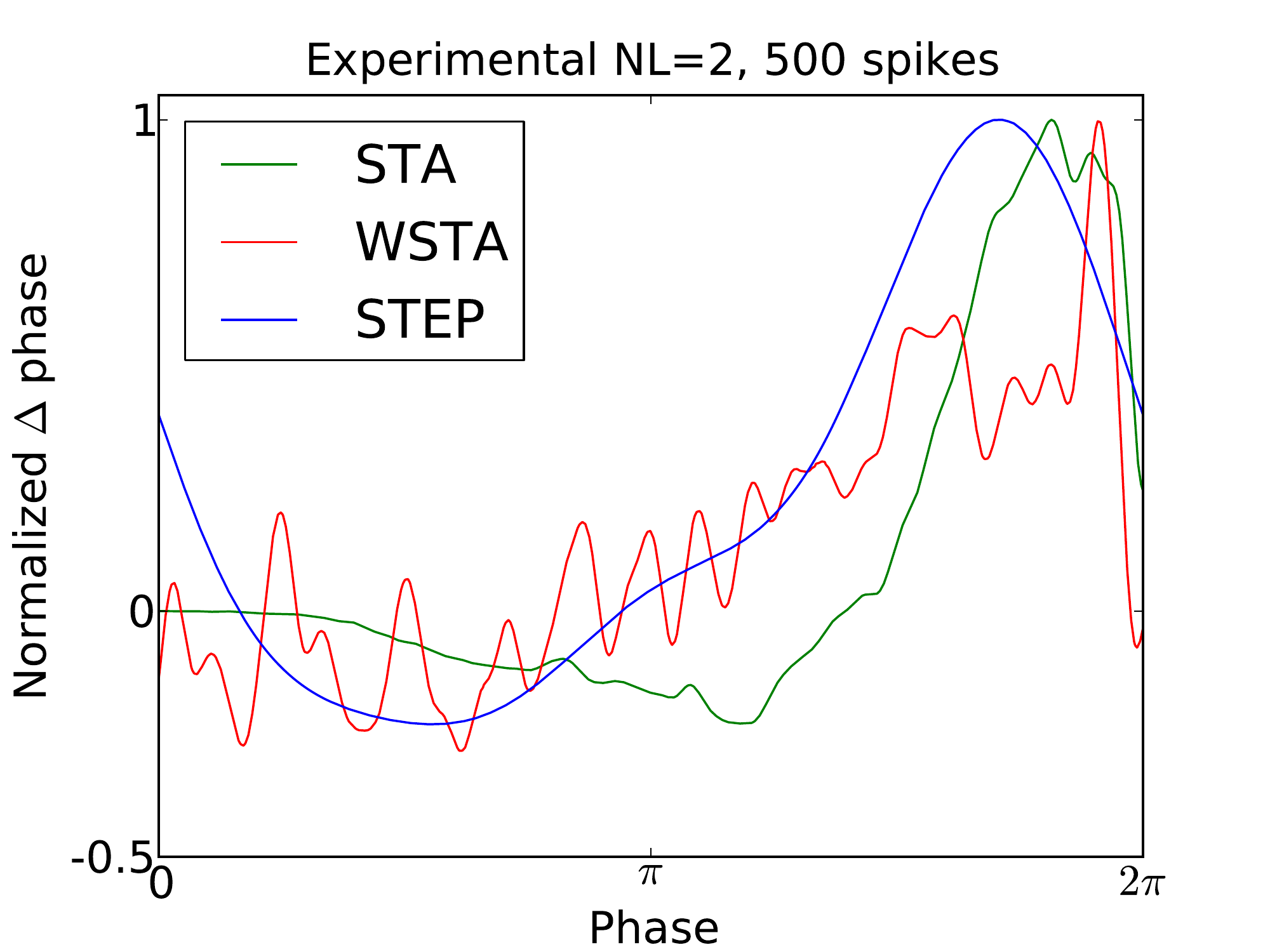} \tabularnewline
\end{tabular}
\caption{The influence of the number of spikes on the reliability of the produced PRCs. The top and bottom row represent noisy modeled data and experimental data, respectively. We tested 50, 100 and 500 spikes and observe that the PRC type is correctly assessed by the STEP and STA method for data sets containing more than 100 spikes. The reliability increases for higher number of spikes but with 500 spikes the STA and STEP methods seem to converge on both modeled data and experimental data while the WSTA method is still very noisy on the experimental data set with 500 spikes.}
\label{fig:nospikes_compare} 
\end{figure}

The reliability of the estimated PRC depends strongly on the number of spikes available for analysis. The data used to obtain the PRCs in Figures~\ref{fig:direct_compare}, ~\ref{fig:noisy_compare}
and~\ref{fig:exp_compare} contain a reasonable number of spikes\footnote{`Reasonable' is here used to denote the number of spikes from experimental data (at least 650) or higher.}, and the results indicate that all three methods are -- to a certain extent -- capable of producing reliable PRCs even under the presence of higher-amplitude fluctuations and more diverse ISIs. However, the number of spikes available for analysis can be limited in experimental data because the experimental protocol is often not exclusively used to gather data to determine a PRC but for other scientific goals. Therefore, we examined the performance of the STA, WSTA and STEP method with less spikes, namely 50, 100 and 500 spikes. Figure~\ref{fig:nospikes_compare} illustrates these results. The columns illustrate the PRC with (from left to right) 50, 100 and 500 spikes (the spikes are the first 50, 100 and 500 of the available spikes in the data set). The PRCs from the top row use spikes from the continuous fluctuating data set at the lowest fluctuation level. The PRCs from the bottom row use experimental data at the second fluctuation level. For the model data, the WSTA and STEP methods give a fair result when using as little as 50 spikes while the STA method produces no usable PRC\footnote{If the spike-triggered average is purely positive, the STA-produced PRC will steeply go negative as illustrated in the top left panel of Figure~\ref{fig:nospikes_compare}.}. With 100 spikes and more, all three methods produce a reliable PRC on this set of model data with little variance in the ISIs.  For the experimental data, a different view emerges. For both 50 and 100 spikes, the WSTA output is useless because of the high noise.  Moreover, the STEP method wrongly classifies the PRC as type-I with only 50 spikes; a negative part in the STEP-produced PRC using 100 or 500 spikes correctly indicates type-II PRC. With 500 spikes, all three method provide a reliable PRC. In contrast to the PRCs from the modeled data, the STA method produces fair PRC with as little as 50 spikes. This result demonstrates that the produced PRCs are influenced by regularity of the data and the number of spikes, rather than claiming superiority of any method of the other. With a different combination of spike times (i.e., not the initial 50, 100, or 500) the results look slightly different (not shown).

The results presented in this paper demonstrate that all methods are capable of producing reliable PRC on pulsed-perturbation data. Moreover, provided a sufficient number of spikes are available, the STA, WSTA and STEP methods also work well with continuous fluctuating data. This result might be, however, misleading since the reliability of the tested methods is sensitive to algorithm parameters such as the a priori estimated ISI ($\widetilde{ISI}$) and the estimated injected step current $I_s$ (on top of which the fluctuations are modulated). These parameters can be set manually or computed by the algorithm itself. Below we describe the effects of these two parameters on the different methods to estimate a PRC.

\begin{figure}[htbp]
 \centering \begin{tabular}{cc}
\includegraphics[scale=0.4]{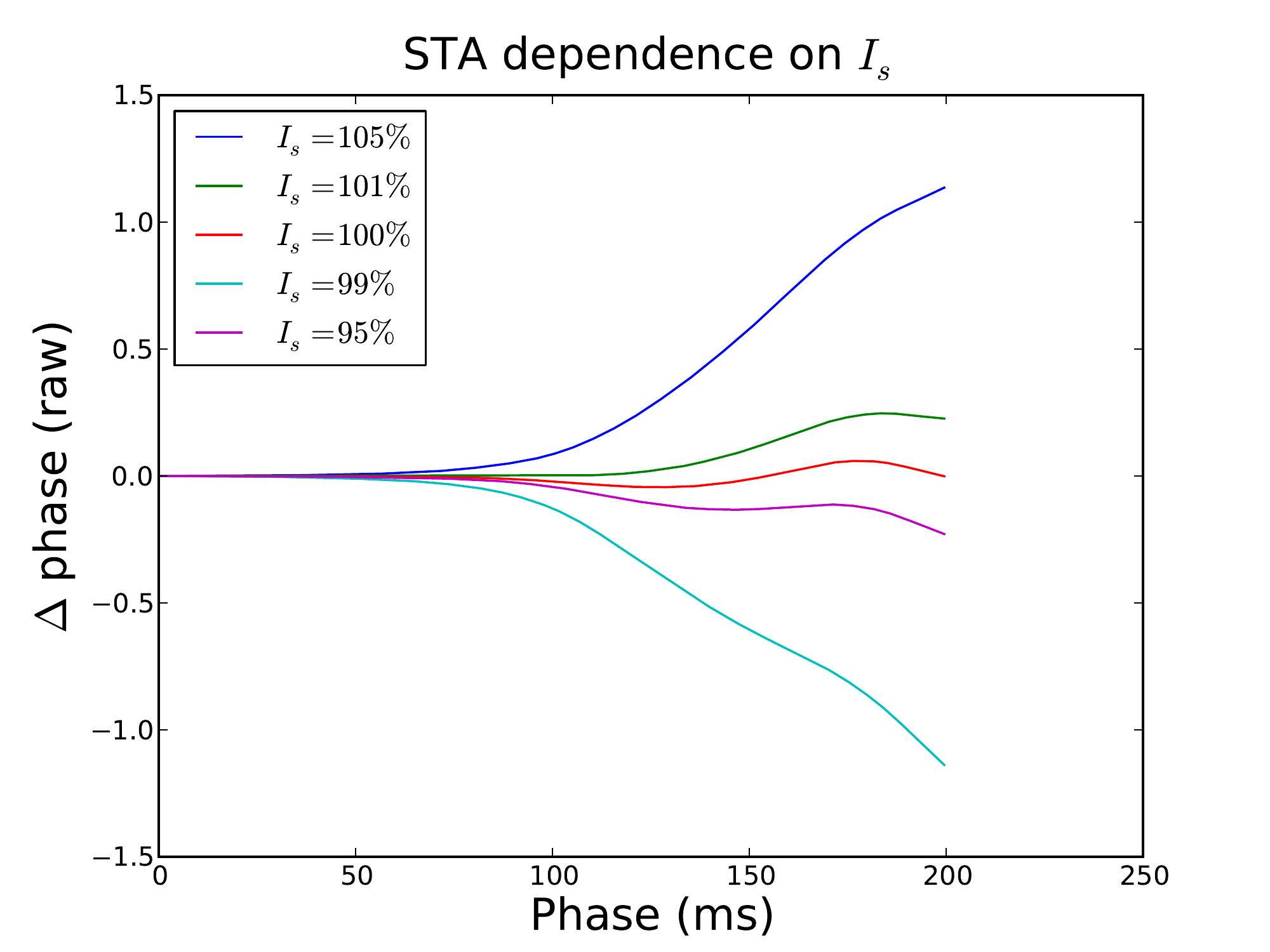}  & 
\includegraphics[scale=0.35]{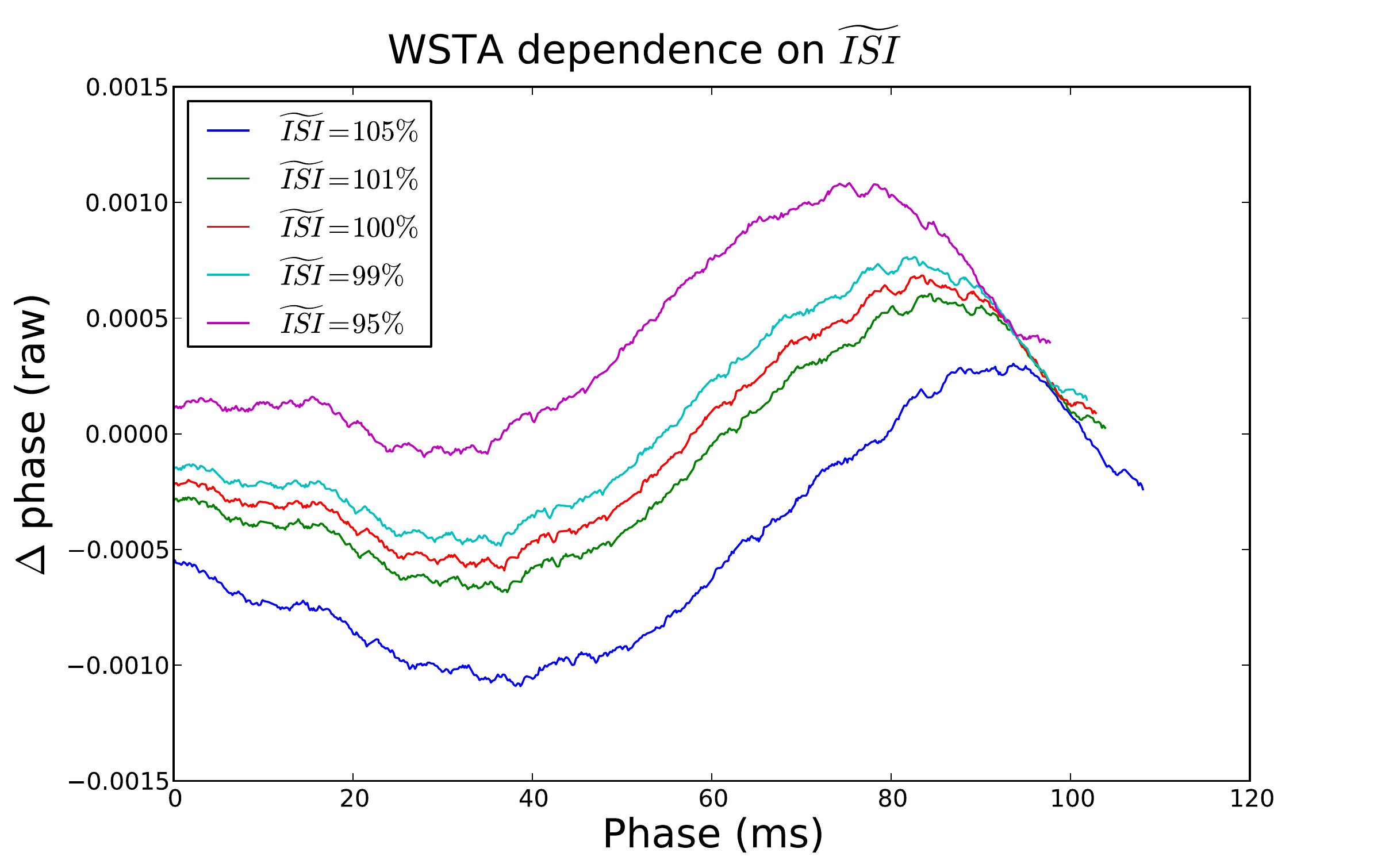} \tabularnewline
\includegraphics[scale=0.4]{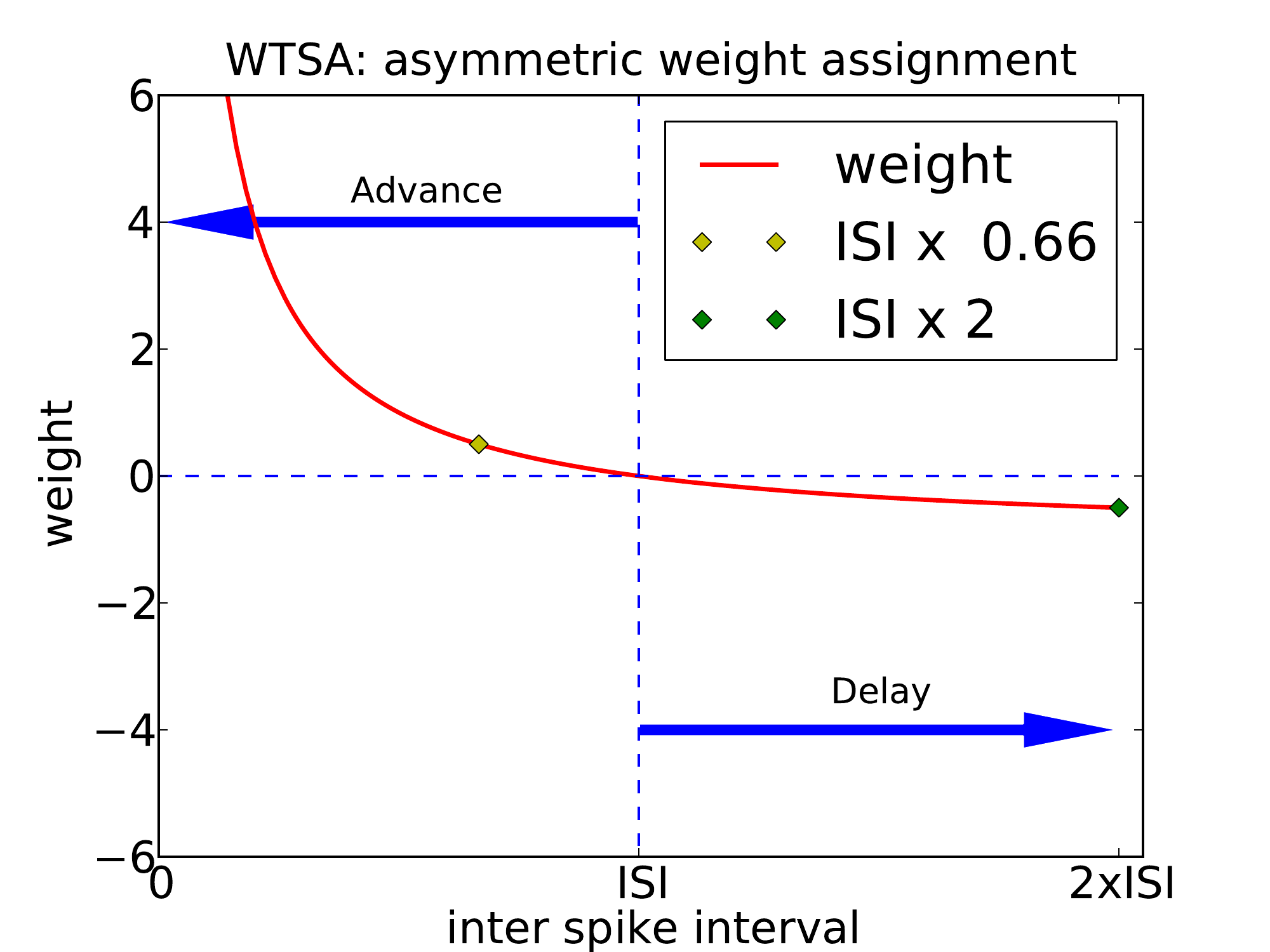}  & 
\includegraphics[scale=0.35]{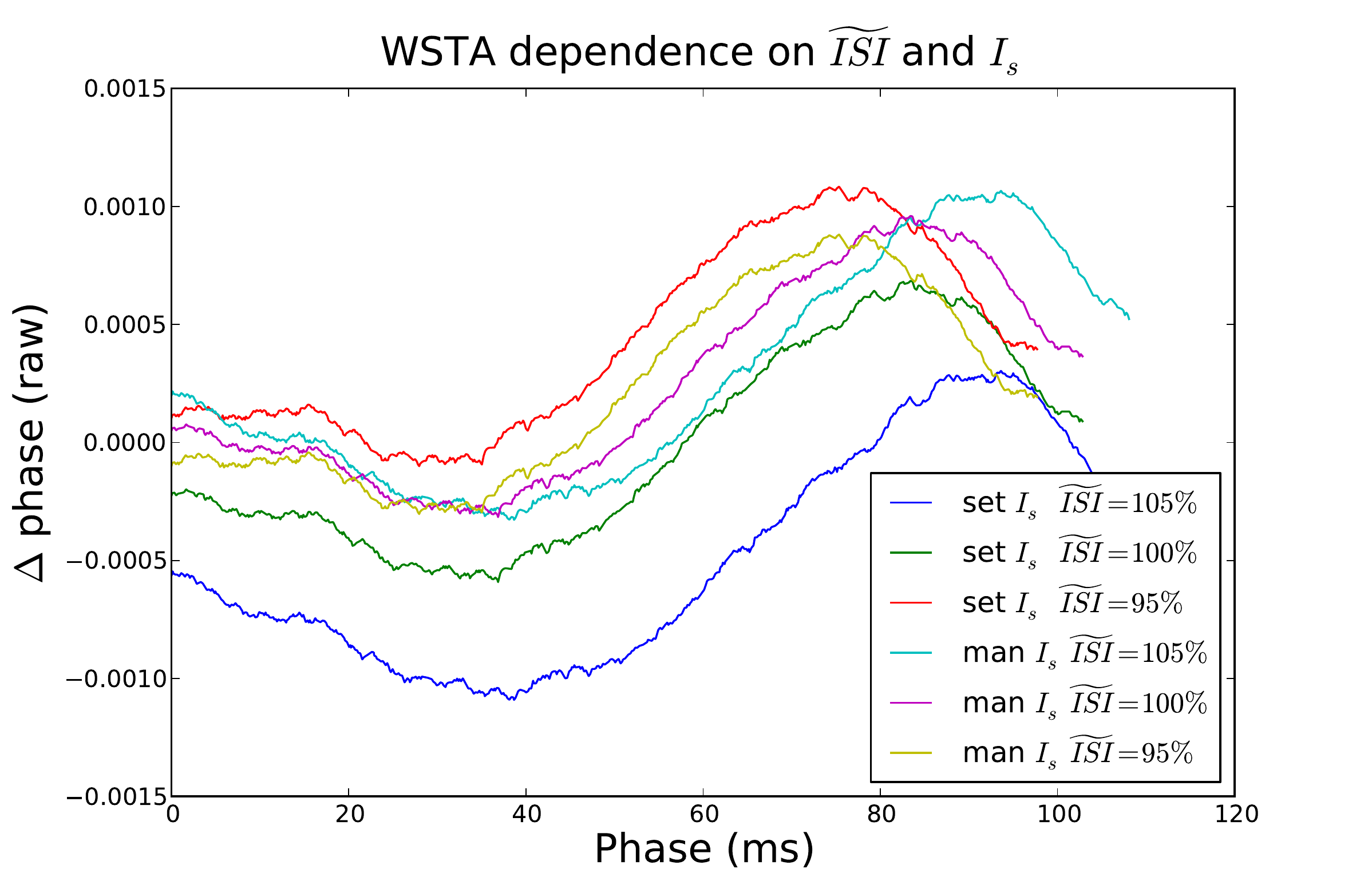} \tabularnewline
\end{tabular}
\caption{Influence of configuration settings in different PRC methods. All PRCs are generated from noisy modeled data with NL=2. Top left: the STA method is highly sensitive to errors in the estimated current step ($I_s$). Top right: the WSTA method is sensitive to estimates of the $\widetilde{ISI}$. Bottom left: the WSTA weighing function is not symmetric and non symmetric distributions of ISIs will lead to upward and downward drifts of the PRC. Bottom right: different outcomes of the WSTA method depend on a combination of the estimated $\widehat{ISI}$ and the step current which can be either manually specified as the predetermined mean (man) or computed from the input fluctuations (set).}
\label{fig:interpretation} 
\end{figure}

We observed that the STA method is highly sensitive to a correct estimation
of the current step, i.e., $I_s$ the injected current to make the neuron fire regularly. Figure~\ref{fig:interpretation}
(top left) illustrates this effect. With a very small deviation of the estimated mean from the real injected DC the outcome becomes unstable. In most cases, this effect will be evident to the researcher: when the amplitude of the spike-triggered average (in the STA method) is close to zero, minor offsets in the estimated DC may shift the spike-triggered average (to either all positive or all negative), and, in turn, shift the PRC resulting in a wrong indication of the PRC type. The top left panel in figure~\ref{fig:nospikes_compare} also clearly demonstrates this effect: the PRC drops steeply (which is accentuated by the normalizing of the positive peak to 1). However, when the spike-triggered average is further away from zero, or when the spike-triggered average crosses zero, a faulty PRC is hard to observe because the resulting PRC will resemble a PRC but will not be representative of the data. Hence, a few different settings for the estimated mean (for instance, the calculated mean from the injected signal in simulations, or, straightforwardly the injected current from the experimental setup) should be used and the resulting PRCs should be compared to PRCs produced by the other methods. In addition, we observed that the PRC produced by the STA is difficult to interpret for two reasons. First the PRC is always shifted on the x-axis towards later phases because it is computed on the interval $[0,ISI_{max}]$ and later normalized to $[0,\widehat{ISI}]$. Second, the y-axis provides the integral of the spike-triggered average; it is not obvious how this relates to the exact delay or advance in spike times. Unfortunately, this second difficulty also arises for the WSTA method where the y-axis is defined by the non-symmetric (see below), weighted sum of the input fluctuations. The STEP method has a straightforward interpretation of the y-axis as it stands for the phase shift.

We observed that the WSTA method is highly sensitive to the estimated ISI (see Figure~\ref{fig:interpretation}, top right). In effect, the $\widetilde{ISI}$ shifts the resulting PRC along the y-axis. This observation can be explained  as follows:  spikes with relatively short ISIs compared to the $\widetilde{ISI}$ (i) are stretched to the $\widetilde{ISI}$ and become straight lines with little variation, and (ii) receive a high weight in the weighted sum (Figure~\ref{fig:interpretation}, bottom left). These two effects combined lead to shorter ISIs pulling the PRC upwards. On the other hand, relatively long ISIs compared to the $\widetilde{ISI}$ make the PRC resulting from the WSTA method noisy because they are compressed to fit on the normalized interval; during this compression smooth fluctuations become steep and are added up to the PRC. For highly regularly firing neurons,
the estimated ISI ($\widetilde{ISI}$) is simply the mean of the ISIs ($\widehat{ISI}$). However, in the
more realistic cases with more variation and a skewed ISI histogram, it becomes less clear what the `best' a priori estimated $\widetilde{ISI}$ should be: mean, median, or mode of the ISIs? In type-II cases this estimation might cause unreliable results because the crossing point (i.e., the point where the curve changes sign) can be shifted along the x-axis and the ratio of positive to negative surface will clearly be modified, even up to the point that the negative part may became insignificant. In addition, figure~\ref{fig:interpretation} (bottom right) illustrates the accumulated effect of different settings for the $\widetilde{ISI}$ and the step current $I_s$. One problem is that all of the PRCs produced by the WSTA method shown in that panel resemble what a researcher might anticipate: some PRC estimates are of type-II while others are of type-I. The difference is caused by changing two configuration options in the algorithm.

In general, the fact that one can configure two settings ($\widetilde{ISI}$ and $I_s$) opens a possibility for a bias in the resulting PRC: as we just illustrated a PRC can be easily `tuned' under the presence of considerable fluctuations to a particular PRC type by changing $\widetilde{ISI}$ and $I_s$. Therefore, one might tune the settings in a way as to prove a particular PRC type. Moreover, even without predispositions about the PRC type, all of the PRC methods (and especially the methods that optimize a smooth curve) can produce PRCs that appear realistic without being representative of the data.

\section{Discussion}

We reviewed five different methods to determine the PRC from experimental
or modeled data. 
Two methods are variations of the direct method and
require the perturbation-stimulation protocol to gather data. The
three other methods use continuous fluctuation data, which requires the
continuous injection of (a step current and) a fluctuation into a regularly firing neuron. We found that on noise-free modeled perturbation data, all methods worked well and showed little difference. Moreover, the two methods requiring pulse perturbation
data produced the best results, but this comes at the cost of a specialized
stimulation protocol and the requirement of a large number of spikes
in order to cover all the phases. In contrast, the methods that
can use the continuous fluctuation data use all available data efficiently
as random fluctuations are by definition delivered at every phase and thus require
less spikes. For instance, one study \cite{galan2005} uses 7000 (highly regular) spikes while we show that the continuous fluctuation methods provide reliable results after a few hundreds spikes
(e.g., 500). Hence, for experimental situations where little time is available it is best to use the continuous fluctuation protocol. When an experiment is done solely for the purpose of determining a PRC, the perturbation protocol should be used.

However, we also demonstrated that the estimated PRC not only depends on the method employed, but also on the settings of the method ($I_s$ and $\widetilde{ISI}$) and the regularity of the spikes (as altered by the amplitude of the fluctuations). Moreover, we demonstrated that the different techniques might generate PRC curves that appear plausible but are not representative for the data. A `panel of experts' strategy can be applied  to enhance the reliability: one can run all the different methods (that are appropriate for the given data set) with slightly different configurations.  A stable PRC for the widest range of settings can be considered the most correct. And, pitfalls such as upward or downward shifts in the PRC can be detected by trying several settings. W also suggest researchers dealing with PRC to inspect carefully the spiking data and obtain good estimates for the $\widetilde{ISI}$ and the DC step current before running the analysis and using any of the PRC estimation methods.

Interpretation of the PRCs is beyond the scope of this review. Briefly, however, different criteria are used to classify PRC curves into type-I curves and type-II curves. For instance, the ratio between the negative amplitude and the positive amplitude \cite{tateno2007} or the ratio between the positive and negative surface \cite{tsubo2007} have been proposed as PRC categorization criteria. Moreover, some reports suggest that the exact shape of the PRC, skewness, zero-crossings and other features contain information about the underlying system, e.g., \cite{gutkin2005,tateno2007}. These features can only be reliably interpreted after obtaining a PRC in a reliable manner following guidelines and avoiding potential pitfalls as outlined above.

We hope that this review of methods for determining the PRC will motivate researchers to determine this measure of spiking behavior for the neural cell types they investigate. The PRC is a measure with considerable predictive power for the behavior of a single neuron in a network  \cite{ermentrout1996}. Given knowledge of a neuron's PRC and its synaptic connections (excitatory/inhibitory, waveform duration, and delay), it is possible to analytically determine its participation in population activities such as synchronization, asynchrony and beating. Specifically, pairs, chains or networks of neurons with a type I PRC will synchronize when coupled by fast inhibitory synapses. For type II neurons, synchrony ensues with excitatory synapses. Strictly speaking, these predictions only hold for homogeneous networks of regularly spiking neurons perturbed by small synaptic potentials. Nevertheless, they allow for insights about network activity emerging from single neuron properties in many activity regimes. Thus, reliable knowledge of PRCs from more neural cell types will aid the understanding of how single neurons contribute to brain function.

\subsubsection*{Acknowledgments}

The authors thank Drs. Yasuhiro Tsubo and Stijn Vanderlooy for fruitful discussions.

\bibliography{bib} 
\bibliographystyle{abbrv}

\end{document}